%% file: main.tex
\documentclass[sigconf,screen]{acmart}

\renewcommand\footnotetextcopyrightpermission[1]{}

\input{preamble}

\AtBeginDocument{%
  \providecommand\BibTeX{{%
    \normalfont B\kern-0.5em{\scshape i\kern-0.25em b}\kern-0.8em\TeX}}}

\begin{document}

\title{\proj: Large Language Models-Aided Program Reduction}

\author{Mengxiao Zhang}
\orcid{0000-0002-3463-2802}
\email{m492zhan@uwaterloo.ca}
\affiliation{%
    \institution{University of Waterloo}
    \city{}
    \country{Canada}
}

\author{Yongqiang Tian}
\authornote{Corresponding author}
\orcid{0000-0003-1644-2965}
\email{yqtian@ust.hk}
\affiliation{%
    \institution{The Hong Kong University of Science and Technology}
    \city{}
    \country{China}
}

\author{Zhenyang Xu}
\orcid{0000-0002-9451-4031}
\email{zhenyang.xu@uwaterloo.ca}
\affiliation{%
    \institution{University of Waterloo}
    \city{}
    \country{Canada}
}

\author{Yiwen Dong}
\orcid{0000-0002-3205-9010}
\email{yiwen.dong@uwaterloo.ca}
\affiliation{%
    \institution{University of Waterloo}
    \city{}
    \country{Canada}
}

\author{Shin Hwei Tan}
\orcid{0000-0001-8633-3372}
\email{shinhwei.tan@concordia.ca}
\affiliation{%
    \institution{Concordia University}
    \city{}
    \country{Canada}
}

\author{Chengnian Sun}
\orcid{0000-0002-0862-2491}
\email{cnsun@uwaterloo.ca}
\affiliation{%
    \institution{University of Waterloo}
    \city{}
    \country{Canada}
}

\begin{abstract}
	\input{abstract}
\end{abstract}




\maketitle

\section{Introduction}
\label{sec:intro}

\input{intro}

\section{Background}
\label{sec:preliminaries}

\input{preliminaries}

\section{Approach}
\label{sec:approach}
\input{approach}

\section{Evaluation}
\label{sec:evaluation}
\input{evaluation}

\section{Discussion}
\label{sec:discussion}
\input{discussion}

\subsection{Threats to Validity}
\label{sec:threats}
\input{threats}

\section{Related Work}
\label{sec:related_work}
\input{related_work}

\section{Conclusion}
\label{sec:conclusion}
\input{conclusion}

\section{Data Availability}
\label{sec:availability}
\input{data_availability}

\input{acknowledgments}

\balance
\bibliographystyle{ACM-Reference-Format}
\bibliography{acmart}

\end{document}

%% file: preamble.tex
\usepackage{standalone}

\usepackage{graphicx}
\usepackage{pdfpages}

\usepackage{colortbl}
\usepackage{xcolor}
\usepackage{array}
\usepackage{makecell}

\definecolor{light}{RGB}{222,235,247}
\definecolor{dark}{RGB}{158,202,225}
\definecolor{darker}{RGB}{49,130,189}
\newcolumntype{a}{>{\columncolor{light}}r}
\newcolumntype{b}{>{\columncolor{dark}}r}
\usepackage[ruled,lined,linesnumbered,vlined]{algorithm2e}

\SetCommentSty{mycommfont}

\usepackage{amsmath}
\usepackage{amsfonts}
\usepackage{amsopn}
\usepackage{amsthm}
\usepackage{nccmath} %
\usepackage{url} %
\usepackage{multicol}
\usepackage{multirow}
\usepackage{setspace} %
\usepackage{xspace}
\usepackage{listings}
\usepackage{stmaryrd} %
\usepackage{todonotes} %
\usepackage{booktabs}
\usepackage{threeparttable}
\usepackage[utf8x]{inputenc}
\usepackage[normalem]{ulem}	%
\usepackage{balance}
\usepackage{subcaption}
\usepackage{tcolorbox}

\usepackage[%
]{hyperref}

\usepackage[scaled=0.87]{beramono}

\usepackage{enumitem}
\usepackage{pdftexcmds}
\usepackage{microtype}
\usepackage{flushend}
\usepackage{pgf}

\newcommand{\proj}{\textsf{LPR}\xspace}

\newcommand{\projPlusVulcan}{\textsf{LPR}+Vulcan\xspace}
\makeatletter

\newcommand{\Rmnum}[1]{\expandafter\@slowromancap\romannumeral
#1@}
\makeatother
\newcommand{\benchmarkSizeAll}{50\xspace}
\newcommand{\benchmarkC}{Benchmark-C\xspace}
\newcommand{\benchmarkSizeC}{20\xspace}
\newcommand{\benchmarkRust}{Benchmark-Rust\xspace}
\newcommand{\benchmarkSizeRust}{20\xspace}
\newcommand{\benchmarkJs}{Benchmark-JS\xspace}

\newcommand{\opFunctioninlining}{\textit{Function Inlining}\xspace}
\newcommand{\opLoopUnrolling}{\textit{Loop Unrolling}\xspace}
\newcommand{\opDataTypeSimplification}{\textit{Data Type Simplification}\xspace}
\newcommand{\opDataTypeElimination}{\textit{Data Type Elimination}\xspace}
\newcommand{\opVariableElimination}{\textit{Variable Elimination}\xspace}

\newcommand{\cSizeAvgPerses}{\cSizeAvgPersesValue\xspace}
\newcommand{\cSizeAvgPersesValue}{247.8}
\newcommand{\cSizeAvgVulcan}{\cSizeAvgVulcanValue\xspace}
\newcommand{\cSizeAvgVulcanValue}{157.4}
\newcommand{\cSizeAvgProj}{\cSizeAvgProjValue\xspace}
\newcommand{\cSizeAvgProjValue}{105.8}
\newcommand{\cSizeStddevProj}{\cSizeStddevProjValue\xspace}
\newcommand{\cSizeStddevProjValue}{4.4}
\newcommand{\cSizeAvgProjPlusVulcan}{\cSizeAvgProjPlusVulcanValue\xspace}
\newcommand{\cSizeAvgProjPlusVulcanValue}{86.1}

\newcommand{\cSizeAvgCreduce}{\cSizeAvgCreduceValue\xspace}
\newcommand{\cSizeAvgCreduceValue}{85.7}

\newcommand{\cMinNumOfSizeProjPlusVulcan}{\cMinNumOfSizeProjPlusVulcanValue\xspace}
\newcommand{\cMinNumOfSizeProjPlusVulcanValue}{13}

\newcommand{\cSizeDecRateVulcanVsPerses}{34.35\%\xspace}
\newcommand{\cSizeDecRateProjVsPerses}{51.33\%\xspace}
\newcommand{\cSizeDecRateProjVsVulcan}{24.93\%\xspace}
\newcommand{\cSizeDecRateProjPlusVulcanVsVulcan}{36.73\%\xspace}

\newcommand{\cSizePValueProjVsVulcan}{0.002\xspace}

\newcommand{\rustSizeAvgPerses}{\rustSizeAvgPersesValue\xspace}
\newcommand{\rustSizeAvgPersesValue}{212.5}
\newcommand{\rustSizeAvgVulcan}{\rustSizeAvgVulcanValue\xspace}
\newcommand{\rustSizeAvgVulcanValue}{184.2}
\newcommand{\rustSizeAvgProj}{\rustSizeAvgProjValue\xspace}
\newcommand{\rustSizeAvgProjValue}{147.5}
\newcommand{\rustSizeAvgProjPlusVulcan}{\rustSizeAvgProjPlusVulcanValue\xspace}
\newcommand{\rustSizeAvgProjPlusVulcanValue}{135.5}

\newcommand{\rustMinNumOfSizeVulcan}{\rustMinNumOfSizeVulcanValue\xspace}
\newcommand{\rustMinNumOfSizeVulcanValue}{9}
\newcommand{\rustMinNumOfSizeProj}{\rustMinNumOfSizeProjValue\xspace}
\newcommand{\rustMinNumOfSizeProjValue}{15}

\newcommand{\rustMinNumOfSizeProjPlusVulcan}{\rustMinNumOfSizeProjPlusVulcanValue\xspace}
\newcommand{\rustMinNumOfSizeProjPlusVulcanValue}{16}

\newcommand{\rustMinNumOfSizeVulcanCaseAvgSize}{\rustMinNumOfSizeVulcanCaseAvgSizeValue\xspace}
\newcommand{\rustMinNumOfSizeVulcanCaseAvgSizeValue}{43}

\newcommand{\rustMinNumOfSizeProjCaseAvgSize}{\rustMinNumOfSizeProjCaseAvgSizeValue\xspace}
\newcommand{\rustMinNumOfSizeProjCaseAvgSizeValue}{269}

\newcommand{\rustSizeDecRateProjVsPerses}{14.87\%\xspace}
\newcommand{\rustSizeDecRateProjVsVulcan}{4.47\%\xspace}
\newcommand{\rustSizeDecRateProjPlusVulcanVsVulcan}{14.39\%\xspace}

\newcommand{\jsSizeAvgPerses}{\jsSizeAvgPersesValue\xspace}
\newcommand{\jsSizeAvgPersesValue}{55.5}
\newcommand{\jsSizeAvgVulcan}{\jsSizeAvgVulcanValue\xspace}
\newcommand{\jsSizeAvgVulcanValue}{38.2}
\newcommand{\jsSizeAvgProj}{\jsSizeAvgProjValue\xspace}
\newcommand{\jsSizeAvgProjValue}{33.9}
\newcommand{\jsSizeAvgProjPlusVulcan}{\jsSizeAvgProjPlusVulcanValue\xspace}
\newcommand{\jsSizeAvgProjPlusVulcanValue}{27.5}

\newcommand{\jsSizeDecRateVulcanVsPerses}{30.35\%\xspace}
\newcommand{\jsSizeDecRateProjVsPerses}{38.66\%\xspace}
\newcommand{\jsSizeDecRateProjPlusVulcanVsVulcan}{28.15\%\xspace}
\newcommand{\jsSizeDecRateProjVsVulcan}{11.71\%\xspace}

\newcommand{\cTimeAvgVulcanValue}{2.146}
\newcommand{\cTimeAvgProj}{\cTimeAvgProjValue\xspace}
\newcommand{\cTimeAvgProjValue}{1.915}

\newcommand{\cTimeFormatAvgVulcan}{2h:08m:46s\xspace}
\newcommand{\cTimeFormatAvgProj}{1h:54m:54s\xspace}

\newcommand{\cTimeDecPercentageOfAvgProjVsVulcan}{%
	\pgfmathparse{(1 - \cTimeAvgProjValue/\cTimeAvgVulcanValue)*100}
    \pgfmathprintnumber{\pgfmathresult}\%\xspace%
}
\newcommand{\cTimeIncAvgOfPercentageProjVsVulcan}{45.83\%\xspace}
\newcommand{\cTimeDecAvgOfPercentageOneHourProjVsVulcan}{4.15\%\xspace}

\newcommand{\cGptTimeAvgProjValue}{0.618}

\newcommand{\cGptTimePercentProj}{%
    \pgfmathparse{(\cGptTimeAvgProjValue/\cTimeAvgProjValue)*100}
    \pgfmathprintnumber{\pgfmathresult}\%\xspace%
}

\newcommand{\rustTimeAvgVulcanValue}{2.824}
\newcommand{\rustTimeAvgProj}{\rustTimeAvgProjValue\xspace}
\newcommand{\rustTimeAvgProjValue}{1.839}

\newcommand{\rustTimeDecPercentageOfAvgProjVsVulcan}{%
	\pgfmathparse{(1 - \rustTimeAvgProjValue/\rustTimeAvgVulcanValue)*100}
    \pgfmathprintnumber{\pgfmathresult}\%\xspace%
}

\newcommand{\rustTimeDecAvgOfPercentageOneHourProjVsVulcan}{21.69\%\xspace}

\newcommand{\rustGptTimeAvgProjValue}{0.517}

\newcommand{\rustGptTimePercentProj}{%
    \pgfmathparse{(\rustGptTimeAvgProjValue/\rustTimeAvgProjValue)*100}
    \pgfmathprintnumber{\pgfmathresult}\%\xspace%
}

\newcommand{\jsTimeAvgVulcanValue}{0.276}
\newcommand{\jsTimeAvgProj}{\jsTimeAvgProjValue\xspace}
\newcommand{\jsTimeAvgProjValue}{0.174}

\newcommand{\jsTimeDecPercentageOfAvgProjVsVulcan}{%
	\pgfmathparse{(1 - \jsTimeAvgProjValue/\jsTimeAvgVulcanValue)*100}
    \pgfmathprintnumber{\pgfmathresult}\%\xspace%
}

\newcommand{\jsGptTimeAvgProjValue}{0.127}

\newcommand{\jsGptTimePercentProj}{%
    \pgfmathparse{(\jsGptTimeAvgProjValue/\jsTimeAvgProjValue)*100}
    \pgfmathprintnumber{\pgfmathresult}\%\xspace%
}

\newcommand{\cSizeChangeFIAlone}{\cSizeChangeFIAloneValue\xspace}
\newcommand{\cSizeChangeFIAloneValue}{-13.2}
\newcommand{\cSizeChangeLUAlone}{\cSizeChangeLUAloneValue\xspace}
\newcommand{\cSizeChangeLUAloneValue}{14.9}
\newcommand{\cSizeChangeDTEAlone}{\cSizeChangeDTEAloneValue\xspace}
\newcommand{\cSizeChangeDTEAloneValue}{-4.7}
\newcommand{\cSizeChangeDTSAlone}{\cSizeChangeDTSAloneValue\xspace}
\newcommand{\cSizeChangeDTSAloneValue}{1.3}
\newcommand{\cSizeChangeVEAlone}{\cSizeChangeVEAloneValue\xspace}
\newcommand{\cSizeChangeVEAloneValue}{-4.3}

\newcommand{\cSizeDecAllFI}{\cSizeDecAllFIValue\xspace}
\newcommand{\cSizeDecAllFIValue}{88.2}
\newcommand{\cSizeDecAllPercentFI}{62.12\%\xspace}

\newcommand{\cSizeDecAllLU}{\cSizeDecAllLUValue\xspace}
\newcommand{\cSizeDecAllLUValue}{4.8}
\newcommand{\cSizeDecAllPercentLU}{3.35\%\xspace}

\newcommand{\cSizeDecFrequencyLU}{6\xspace}

\newcommand{\cSizeAvgSingleLevelProj}{\cSizeAvgSingleLevelProjValue\xspace}
\newcommand{\cSizeAvgSingleLevelProjValue}{155.0}
\newcommand{\cSizeStddevSingleLevelProj}{\cSizeStddevSingleLevelProjValue\xspace}
\newcommand{\cSizeStddevSingleLevelProjValue}{8.7}

\newcommand{\gcc}{GCC\xspace}
\newcommand{\llvm}{LLVM\xspace}
\newcommand{\clang}{Clang\xspace}
\newcommand{\rust}{Rust\xspace}

\newcommand{\javascript}{JavaScript\xspace}
\newcommand{\javascriptcore}{JavaScriptCore\xspace}
\newcommand{\js}{JS\xspace}
\newcommand{\java}{Java\xspace}
\newcommand{\smt}{SMT-LIBv2\xspace}

\newcommand{\libtooling}{LibTooling\xspace}
\newcommand{\candcpp}{C\xspace}

\newcommand{\creduce}{C-Reduce\xspace}
\newcommand{\jreduce}{J-Reduce\xspace}
\newcommand{\perses}{Perses\xspace}

\newcommand{\hdd}{HDD\xspace}

\newcommand{\hierarchicaldeltadebugging}{Hierarchical
 Delta Debugging\xspace}
\newcommand{\ddmin}{DDMin\xspace}
\newcommand{\ddsmt}{ddSMT\xspace}

\newcommand{\llms}{LLMs\xspace}
\newcommand{\llm}{LLM\xspace}
\newcommand{\llmsfull}{Large Language Models\xspace}
\newcommand{\chatgpt}{ChatGPT\xspace}
\newcommand{\gptCurrentVersion}{gpt-3.5-turbo-0613\xspace}

\newcommand{\llama}{CodeLlama\xspace}
\newcommand{\llamaVersiona}{CodeLlama-13b\xspace}
\newcommand{\llamaVersionb}{CodeLlama-34b\xspace}
\newcommand{\openaiapi}{OpenAI API\xspace}
\newcommand{\vulcan}{Vulcan\xspace}
\newcommand{\apr}{APR\xspace}

\newcommand{\program}{\ensuremath{P}\xspace}
\newcommand{\property}{\ensuremath{\psi}\xspace}
\newcommand{\programmin}{\ensuremath{\program_\textit{min}}\xspace}
\newcommand{\programtmp}{\ensuremath{\program_\textit{tmp}}\xspace}
\newcommand{\transformationListMath}{\ensuremath{\textit{transformList}}\xspace}
\newcommand{\transformationMath}{\textit{transform}\xspace}
\newcommand{\promptMath}{\textit{prompts}\xspace}
\newcommand{\targetListMath}{\ensuremath{\textit{targetList}}\xspace}
\newcommand{\targetMath}{\textit{target}\xspace}
\newcommand{\primaryQuestionMath}{\ensuremath{\textit{primaryQuestion}}\xspace}
\newcommand{\followupQuestionMath}{\ensuremath{\textit{followupQuestion}}\xspace}

\newcommand{\searchspace}{\ensuremath{\mathbb{P}}\xspace}

\newcommand{\boolspace}{\ensuremath{\mathbb{B}}\xspace}

\newcommand{\pvalue}{p-value\xspace}

\newcommand{\aka}{\hbox{\emph{a.k.a.}}\xspace}

\newcommand{\etal}{\hbox{\emph{et al.}}\xspace}

\newcommand{\eg}{\hbox{\emph{e.g.}}\xspace}
\newcommand{\ie}{\hbox{\emph{i.e.}}\xspace}

\newcommand{\wrt}{\hbox{\emph{w.r.t.}}\xspace}

\SetKwFunction{AlgReduce}{Reduce}
\SetKwFunction{AlgRemove}{Remove}
\SetKwFunction{AlgMinSubtree}{MinSubtree}
\SetKwFunction{AlgUndoReplace}{UndoReplace}
\SetKwFunction{AlgInsert}{Insert}
\SetKwFunction{AlgDelete}{Delete}
\SetKwFunction{AlgReplace}{Replace}
\SetKwFunction{AlgMain}{\proj}
\SetKwFunction{AlgPatch}{Patch}
\SetKwFunction{AlgTreeDiff}{$\Delta_T$}
\SetKwFunction{AlgListDiff}{$\Delta_L$}
\SetKwFunction{AlgPop}{Pop}
\SetKwFunction{AlgPush}{Push}
\SetKwFunction{AlgRemainingNodes}{GetRemainingNodes}
\SetKwFunction{AlgReducePlus}{ReducePlus}
\SetKwFunction{AlgReduceOptional}{ReduceOptional}
\SetKwFunction{AlgReduceRegular}{ReduceRegular}
\SetKwFunction{AlgGetAndRemoveLargestFrom}{GetAndRemoveLargestFrom}
\SetKwFunction{AlgChildren}{Children}
\SetKwFunction{AlgParent}{Parent}
\SetKwFunction{AlgDdmin}{ddmin}
\SetKwFunction{AlgDd}{dd}
\SetKwFunction{AlgParse}{ParseTree}
\SetKwFunction{AlgCopy}{Duplicate}
\SetKwFunction{AlgGetTransformationList}{getTransformList}
\SetKwFunction{AlgGetTargetList}{getTargetList}
\SetKwFunction{AlgApplyTransformation}{applyTransformation}
\SetKwFunction{AlgGetPrimaryQuestion}{getPrimaryQuestion}
\SetKwFunction{AlgGetFollowupQuestion}{getFollowupQuestion}
\SetKwFunction{AlgPerses}{Perses}

\SetKwData{AlgMin}{Min}
\SetKwData{AlgMinIndex}{min\_index}
\SetKwData{AlgResult}{result}

\SetKwFunction{Encode}{Encode}
\SetKwFunction{CompactEncode}{CompactEncode}
\SetKwFunction{Decode}{CompactDecode}
\SetKwFunction{Refresh}{RefreshCache}

\SetKwFunction{BoundedBFS}{BoundedBFS}
\SetKwFunction{IsKleene}{IsKleene}
\SetKwFunction{RuleType}{Rule}
\SetKwFunction{ExpectedRuleType}{ExpectedRule}
\SetKwFunction{QuantifiedRuleType}{QuantifiedRule}

\SetKwData{Candidates}{candidates}
\SetKwData{Predicate}{pred}

\SetKwData{AlgTrueLiteral}{true}
\SetKwData{AlgFalseLiteral}{false}
\SetKwData{AlgMin}{min}
\SetKwData{AlgCache}{cache}
\SetKwData{AlgPrev}{prev}
\SetKwData{AlgWorklist}{worklist}
\SetKwData{AlgNext}{pending}
\SetKwData{AlgLargest}{largest}

\SetKw{Continue}{continue}

\SetKwProg{Fn}{Function}{:}{}

\newcommand{\mycode}[1]{\texttt{#1}\xspace}

\newcommand{\HighlightBlack}[1]{}

\newcommand{\myparagraph}[1]{
	\vspace*{0.04cm}
	\noindent \textit{\textbf{#1.}}\quad
}

\usepackage{tikz}
\newcommand*\circled[1]{\tikz[baseline=(char.base)]{
		\node[shape=circle,draw,inner sep=1pt] (char) {#1};}}

\usepackage{tikz}
\usepackage{forest}

\definecolor{deepRed}{HTML}{D81B60}
\definecolor{deepBlue}{HTML}{1E88E5}
\definecolor{deepOrange}{HTML}{FFC107}

\newcommand{\verticalline}{\unskip\ \vrule\ \ \ }

\newcounter{num}

\newcommand{\finding}[1]{
	\begin{tcolorbox}[arc=4pt,left=3pt,right=3pt,top=3pt,bottom=3pt,boxsep=0pt]
		\textbf{RQ\refstepcounter{num}\thenum}: #1
	\end{tcolorbox}
}

\usepackage{cleveref} %

\Crefname{table}{Table}{Tables}
\crefname{table}{Table}{Tables}
\Crefname{figure}{Figure}{Figures}
\crefname{figure}{Figure}{Figures}
\Crefname{algocf}{Algorithm}{Algorithms}
\crefname{algocf}{Algorithm}{Algorithms}
\Crefname{algorithm}{Algorithm}{Algorithms}
\crefname{algorithm}{Algorithm}{Algorithms}
\crefname{thm}{Theorem}{Theorems}
\Crefname{thm}{Theorem}{Theorems}
\crefname{appendix}{Appendix}{Appendices}
\Crefname{appendix}{Appendix}{Appendices}

\crefformat{chapter}{\S#2#1#3}
\crefmultiformat{chapter}{\S\S#2#1#3}{ and~#2#1#3}{, #2#1#3}{, and~#2#1#3}

\crefformat{section}{\S#2#1#3}
\crefmultiformat{section}{\S\S#2#1#3}{ and~#2#1#3}{, #2#1#3}{, and~#2#1#3}

%% file: abstract.tex
Program reduction is a widely used technique to facilitate
debugging compilers by automatically
minimizing programs that trigger compiler bugs.
Existing program reduction
techniques are either generic to a wide range of languages
(such as \perses and
\vulcan) or specifically optimized for one certain language by exploiting language-specific
knowledge (\eg, \creduce).
However, synergistically combining both generality
across languages and optimality to a specific language in program
reduction is yet to be explored.

This paper proposes \proj,
the first \llms-aided technique leveraging \llms
to perform language-specific program reduction
for multiple languages.
The key insight
is to utilize both the
language generality of program reducers such as \perses
and the language-specific semantics learned by \llms.
Concretely, language-generic program reducers can
efficiently reduce programs into
a small size that is suitable for \llms to process;
\llms can effectively transform
programs via the learned semantics to create
new reduction opportunities for the language-generic program
reducers to further reduce the programs.

Our thorough evaluation on \benchmarkSizeAll
benchmarks across three programming
languages (\ie, \candcpp, \rust and \javascript) has demonstrated
\proj's practicality and superiority over \vulcan, the state-of-the-art
language-generic program reducer.
For effectiveness, \proj
surpasses \vulcan by
producing \cSizeDecRateProjVsVulcan, \rustSizeDecRateProjVsVulcan,
and \jsSizeDecRateProjVsVulcan smaller
programs on benchmarks in C, \rust and
\javascript, separately.
Moreover, \proj and \vulcan have the potential to
complement each other. For the \candcpp language
for which \creduce is optimized,
by applying \vulcan to the
output produced by \proj, we can attain program
sizes that are on
par with those achieved by \creduce.
For efficiency perceived by users,
\proj is more efficient when reducing large and
complex programs, taking
\cTimeDecPercentageOfAvgProjVsVulcan,
\rustTimeDecPercentageOfAvgProjVsVulcan,
\jsTimeDecPercentageOfAvgProjVsVulcan less time
than \vulcan
to finish all the benchmarks in C,
\rust and \javascript, separately.

%% file: intro.tex
Program reduction
techniques~\cite{sun2018perses, xu2023pushing, regehr2012test,
kalhauge2019binary, kalhauge2021logical, niemetz2013ddsmt,
rcc, wang2021probabilistic, misherghi2006hdd,
donaldson2021test}
aim to
facilitate compiler debugging by minimizing the bug-triggering programs
with efficacy and efficiency.
Given a program \program and a property \property that \program
preserves, program reduction techniques (\aka, program reducers)
produce a minimal
program
\programmin that still preserves \property.
Program reduction
has been widely used in various software engineering tasks~\cite{sigplan,chisel},
especially in compiler testing and debugging~\cite{livinskii2020random,ccmd,
ProgramReconditioning,kitten,adhoc}.

However, a critical challenge in program reduction
has not been properly addressed,
\ie, the trade-off between generality
across languages and specificity to a certain language.
Currently, there are two categories of program reduction techniques:
language-specific
reduction~\cite{regehr2012test,kalhauge2019binary,kalhauge2021logical}
and
language-generic
reduction~\cite{zeller2002simplifying,misherghi2006hdd,sun2018perses,xu2023pushing,zhang2023ppr}.
The former category leverages language-specific semantics to transform
and shrink programs in certain languages,
while the latter
only uses transformations
applicable to any programming language.
Although language-specific reducers are usually more effective in
reduction,
designing an effective
reducer for a specific
language,
especially language-specific transformations,
requires a deep understanding of language features
and a significant amount of time and engineering effort.
Therefore,
only a limited set of  languages
have custom-designed reducers,
such as \candcpp~\cite{regehr2012test}, \java~\cite{kalhauge2019binary,
    kalhauge2021logical}, and \smt~\cite{niemetz2013ddsmt}.
Meanwhile, language-generic reducers can be applied to diverse languages,
but
lack the knowledge of language features and semantics
and thus are incapable of performing language-specific transformations
(\eg, function inlining) that can enable further reduction.
As a result, they cannot utilize
peculiar features of each language to produce optimally reduced programs.

This study strives to find a sweet spot
between generality
across languages and specificity to a certain language,
by synergistically combining the strengths of both categories of program reduction techniques.
Specifically,
we notice that
the major limitation of language-generic reducers lies in their incapability
to perform language-specific transformations.
Language-generic reducers such as \perses stand out as high generality
when reducing programs across various programming languages, while
they lack awareness of semantic information to achieve further progress.
If we could help language-generic reducers
conquer this limitation,
they are likely to produce smaller results.

Meanwhile, we also notice that
recent \llmsfull (\llms)
could be powerful assistants in
performing language-specific transformations,
like how they perform in other
software engineering tasks such as
code generation and test generation~\cite{smtsolverllm2023sun,synthesisllm2022jain,copiloting2023wei,
deng2023largezeroshot,fan2023automated,xia2023automated,llmcodegen2023gu}.
Specifically,
\llms are trained with massive programs,
and they have started to exhibit the ability
to analyze and transform programs in prevalent
languages.
If we can properly leverage this ability for program reduction,
we may have a language-generic reducer
being aware of the semantics of
various prevalent languages.
In addition, the utilization of \llms can
simplify the customization and extension of
reducers, as it would be time-consuming and labor-intensive to manually implement
a language-specific reducer or add functionality to an existing one
(such as \creduce using the \clang frontend to implement C/C++-specific
program transformations).

\myparagraph{Challenges of Using \llms}
\llms are not the silver bullet
to program reduction.
When \llms are instructed to handle source code as inputs,
due to the intrinsic limitations of
\llms~\cite{fan2023large,wang2024software,li2023halueval},
\llms may be unable to understand
subtle differences in code~\cite{li2023finding}
and be distracted by
irrelevant
context~\cite{shi2023large}.
Moreover,
in program reduction,
input programs typically contain tens of
thousands of line of code,
which surpasses the input limits of \llms.
Besides, without
effective guidance, \llms are unclear
about what transformations to perform.

\myparagraph{\llms-Aided Program Reduction (\proj)}
We propose \proj in this paper, which is,
to the best of our knowledge,
the first approach that integrates \llms for
program reduction tasks.
\proj synergistically leverages the strengths of
both language-generic program reducers and \llms.
Specifically,
\proj alternates between invoking a language-generic reducer (we use \perses in
experiments)
and the \llm. Initially, \perses efficiently reduces large programs to a
size manageable for the \llm. Subsequently, the \llm further transforms \perses's
output based on specific user-defined prompts that dictate the required
transformations. Following this, \perses is re-invoked, as
transformations made by the \llm often create additional opportunities for
reduction. This process iterates until
the program cannot be further minimized.
For transformations, we have identified five
language-generic transformations to enable further reduction:
\opFunctioninlining, \opLoopUnrolling,
\opDataTypeElimination, \opDataTypeSimplification, and
\opVariableElimination.

To address the aforementioned challenge of using \llms,
\proj is designed with a multi-level prompting approach.
In detail, \proj initially requests the \llm to
identify a list of potential targets for a given transformation, and then
sequentially instructs the \llm to apply the transformation on each target. The
multi-level
prompt guides the \llm in a more concentrated way, by excluding irrelevant
context and other targets that may distract the \llm.

We have conducted extensive evaluations on \proj, illustrating its superiority
over \vulcan, the state-of-the-art language-generic algorithm. On three
benchmark suites, \ie, \benchmarkC, \benchmarkRust and \benchmarkJs,
\proj produces significantly smaller programs than \vulcan by
\cSizeDecRateProjVsVulcan, \rustSizeDecRateProjVsVulcan and
\jsSizeDecRateProjVsVulcan, separately.
Moreover, \proj and \vulcan complement each other.
For \candcpp language
which \creduce is optimized for,
by applying \vulcan to the
output produced by \proj, we attain program
sizes that are on
par with those achieved by \creduce.
For reduction efficiency,
\proj performs
comparably to \vulcan.
In terms of execution time perceived by users, \proj is
more efficient than \vulcan on reducing complex programs. Furthermore, our
detailed analysis
indicates that each of the proposed transformations plays a crucial role in the
reduction process.
We also compare the \proj's performance with the multi-level
prompt
against that without
it, illustrating the efficacy of our proposed multi-level prompting
approach.

\myparagraph{Contribution}
This study makes the following contributions.
\begin{itemize}[leftmargin=*]
	\item We introduce \proj, the
	first attempt to use \llms for program reduction.
	By synergizing the capabilities of both language-generic program reducers and \llms,
	\proj exhibits both generality across various languages
	and awareness of semantics in specific languages. Moreover,
	\proj is easy and flexible to extend with new transformations by simply
	designing new prompts.

	\item We propose a multi-level prompting approach to guide \llms to execute program
	transformations, and demonstrate its effectiveness in practice.
	We propose five general-purposed transformations for \llms to
	reduce
	programs or create more reduction opportunities for language-generic program reducers to exploit.

	\item We comprehensively evaluated \proj on \benchmarkSizeAll benchmarks across
	three
	commonly used languages: \candcpp, \rust and \javascript.
	Results demonstrate \proj's strong effectiveness
    and
	generality.

\end{itemize}

%% file: preliminaries.tex
\subsection{Program Reduction}
Given a program \program with a certain property, \eg, triggering
a compiler bug,
the goal of program reduction is to search for a minimal program
\programmin that still triggers the bug. Program reduction
has demonstrated its significant usefulness in removing bug-irrelevant
code snippets.
The original bug-triggering code~\cite{yang2011finding, le2014compiler,
sun2016finding, livinskii2020random} may have thousands of lines,
whereas the
distilled version from program reduction tools only
contains dozens of lines of code~\cite{sun2016toward}.
Some algorithms are designed to generalize across multiple programming
languages, while others are customized for certain languages.

\subsubsection{Language-Generic Reducers}
Some reducers can be generalized to multiple languages. For instance,
given the formal syntax of a programming language, algorithms like \hdd
and \perses can be used to reduce programs corresponding to this
language. \hdd parses the program into a parse tree and then applies the
\ddmin~\cite{zeller2002simplifying} at each level of the tree to remove
unnecessary tree
nodes as much as possible. \perses goes further than \hdd by performing
certain transformations on the formal syntax to avoid generating
syntactically incorrect program variants.
\vulcan extends \perses, by introducing
novel auxiliary reducers to
exhaustively search for smaller variants by replacing
identifiers/sub-trees and deleting local combinations of tree
nodes on the parse tree.

However, different languages possess their own unique semantic features.
Although the aforementioned algorithms are relatively
efficient~\cite{sun2018perses},
they are incapable of utilizing unique semantics of a particular language to
further reduce a program. For example, these algorithms lacks the ability to
perform transformations like function inlining. Although \vulcan can identify
more reduction opportunities through transformations such as identifier
replacement and local exhaustive search, its approach is akin to "brute
force" enumeration. This method lacks awareness of the given program's
semantics, making it less effective and efficient overall.

\subsubsection{Language-Specific Reducers}
Previous work has introduced reducers customized for
some specific languages. For example, \creduce~\cite{regehr2012test} is
the most effective
reduction tool for \candcpp code. It comprises multiple passes that
transform
the program based on features of the language,
thereby making it
smaller. Language-specific reducers often rely on static program analysis
tools for analysis and modification, \eg,
\libtooling~\cite{libtooling} is employed in \creduce.

However, developing a language-specific reducer
is nontrivial.
To the best of our
knowledge, only a few languages have specific reducers, such as
\candcpp~\cite{regehr2012test},
\java~\cite{kalhauge2019binary, kalhauge2021logical}, and
\smt~\cite{niemetz2013ddsmt}.
The reason is
that the process of
designing a reducer for
a language, or adding
new transformations to an existing reducer, is time-consuming and
labor-intensive.
For instance, in version 2.10.0 of
\creduce~\cite{creducegithub}, function inlining was implemented with 604
lines of C++ code. Such challenges impede the development and
maintenance of language-specific reducers.

\subsection{\llmsfull}

 \llmsfull refer to a type of deep learning
 models trained on huge
 data sets for diverse tasks.
The advent of \llms has opened up
numerous potential opportunities across diverse research fields. \llms are
not only proficient in processing natural languages but also exhibit
substantial capabilities in understanding and processing programming
languages. This highlights the promising future and evolutionary prospects
in the realm of software engineering. Recently, \llms have been applied and
assessed on various software engineering tasks, such as automatic
program repair~\cite{huang2023empirical, xia2023keep,
	xia2023automated, xia2023revisiting,fan2023automated} and program
generation~\cite{liu2023your, zhong2023study, tian2023chatgpt}.

However, despite the usefulness of \llms, some
researchers~\cite{li2023finding}
illustrate
that current \llms are weak in distinguishing nuances between programs.
Moreover, the
memorizing and processing capacity of \llms deteriorates as the input size
grows~\cite{fan2023large,shaham2022scrolls}.
Besides, one cannot expect
\llms to
automatically complete complex tasks; they must be guided
accordingly~\cite{xia2023automated,li2023finding}.
Therefore, for program reduction, directly asking \llms to reduce programs
with tens of thousands of lines is impractical.

%% file: approach.tex
In this section, we first introduce a motivating example, and then
we provide an
overview of the \proj workflow.
We also outline the
details of prompts and proposed transformations in the workflow
that enable the \llm to function effectively on given programs.

\subsection{Motivation}
\input{figures/code_llvm_31259_loop_unrolling}
A motivating example is displayed
in~\cref{fig:code_llvm_31259_loop_unrolling}, where
the original code contains highly nested loops, shown
in~\cref{subfig:loop_unrolling:original}. From
\cref{subfig:loop_unrolling:before_op} to
\cref{subfig:loop_unrolling:after_op}, the nested loops are fully unrolled into
hundreds of lines via \opLoopUnrolling, based on the semantic transformations from
the \llm. Despite the temporary size increase, the
following \perses effectively eliminates all lines
except for the bug-relevant one. This is also the final result of \proj, presented
in~\cref{subfig:loop_unrolling:after_perses}.
By contrast, as shown in ~\cref{subfig:loop_unrolling:vulcan}, \vulcan is incapable of
escaping the local minima by exhaustively replacing identifiers and tree
nodes.
Similarly, \creduce
cannot fully break down the loop structures, given that it is not
integrated with transformations to
unroll loops. Even though loop unrolling techniques
can be added into \creduce in future versions, it will be labor-intensive to
implement a specific transformation compared to user-defined prompts in
natural language.

\subsection{Workflow}

\input{plots/workflow}

\input{alg/workflow}

The overview of the workflow is outlined in~\cref{fig:workflow}.
Given a bug-triggering program as input,
\proj invokes a language-generic reducer and the \llm alternately, until the
target program cannot be further reduced.
In each iteration, the
language-generic reducer
efficiently reduces the given program to
a minimal one. By
contrast, the \llm leverages the
semantic knowledge of the language and transforms the program
given by
the language-generic reducer.
This process is guided by user-defined prompts,
aiming to expose more reduction potentials to the
language-generic reducer.

~\cref{alg:main}
shows \proj's reduction algorithm.
Given as inputs (1) a program \program targeted for reduction, (2) a
property
\property
that must be preserved, and (3) the pre-defined
\promptMath,
\proj generates a reduced program
$P_{min}$.
The pre-defined
\promptMath consist of \primaryQuestionMath and
\followupQuestionMath,
in which \primaryQuestionMath instructs the \llm to identify a list of targets
to be
transformed,
and \followupQuestionMath guides the \llm to apply transformation on each target
individually.
They will be further introduced
in~\cref{subsection:prompts}.
Initially,
\proj calls \perses to quickly minimize \program, so that
the size of the program becomes manageable for the \llm to process.
Next,
\proj loads a sequence of transformations as delineated in
\cref{alg:main:get_transformation_list},
and then iterates through each transformation,
as detailed from \cref{alg:main:loop_transformation_start} to
\cref{alg:main:loop_transformation_end}. \cref{subsection:transformations} displays
the details of each transformation.

During this process, for each transformation, the algorithm retrieves a
predefined primary question along with a follow-up question
on~\cref{alg:main:get_primary_question} -- \cref{alg:main:get_followup_question}.
\proj firstly asks the \llm the primary question under the current
program.
This
query aims to guide the \llm to generate a list of specific
targets upon which the transformation will be executed. For instance, for
\opLoopUnrolling in the motivating example
~\cref{subfig:loop_unrolling:before_op}, the \llm is asked to identify a list of loops in
the given program to be unrolled in the \primaryQuestionMath, and returns a target
list $[ \mycode{for (i = 0; ...)},
\mycode{for ( j = 0; ...)},$
$\mycode{for ( k = 0; ...)}]$.

On \cref{alg:main:loop_target_start} to
\ref{alg:main:loop_target_end}, \proj uses \followupQuestionMath to guide
the \llm to apply the
transformation on each identified target within the program. In the motivating
example, the \followupQuestionMath can be framed as ``Given the program
\{ PROGRAM \} and the loop \mycode{for(i=0;...)}, optimize it via loop
unrolling''.
 The modified program is then extracted from the \llm's response text
on~\cref{alg:main:perform_transformation}. In the example, all loops are
unrolled into repeated lines of code in ~\cref{subfig:loop_unrolling:after_op}.

In experimental scenarios, given an input program and the prompt, the \llm may
generate multiple transformed programs, as the number
of responses can be customized.
Among all transformed programs, \proj keeps the smallest one that still
passes the
property test, and discards
 others. If no transformed program returned from this query satisfies the property,
 \proj keeps the original one before this query.
 After each transformation is completed, a language-generic reducer such as \perses~\cite{sun2018perses}
 is
employed to seek additional reduction opportunities, considering that the
transformation might have introduced new potential for further
simplification.
The algorithm persists in the outermost loop until it reaches a fixpoint,
signifying that the program size can no longer be reduced.

\input{plots/prompts}
\subsection{Transformations}
\label{subsection:transformations}
To further search for reduction opportunities on
a bug-triggering program via the \llm, we
propose five general transformations to guide the \llm, \ie, \opFunctioninlining,
\opLoopUnrolling,
\opDataTypeElimination, \opDataTypeSimplification and
\opVariableElimination.

\myparagraph{\opFunctioninlining}
This transformation identifies a function and inlines it to eliminate
all call sites of this function, and instead
substitutes them with the corresponding function body. As
functions are commonly found in bug-triggering
programs,
there is significant room for function inlining to reduce tokens or provide
further reduction opportunities.

\myparagraph{\opLoopUnrolling}
Loop unrolling, also known as loop unwinding, is a widely used loop
transformation approach to optimize the execution. In this task, it can also
be employed to find more reduction opportunities.
\opLoopUnrolling identifies a loop structure and attempts to unroll the loop
into a code snippet repeating a single iteration.
This is motivated by the fact that
language-generic reducers may be incapable of dismantling or directly removing
the loop structure, while they may be able to reduce the repeated code after loop
unrolling, as shown in ~\cref{fig:code_llvm_31259_loop_unrolling}.

\myparagraph{\opDataTypeElimination}
Some data types in bug-triggering programs may be irrelevant to the bug,
such as identifiers defined by
\mycode{typedef} in \candcpp, and type alias created by \mycode{type}
keyword in \rust.
We propose \opDataTypeElimination to
eliminate the alias and replace the occurrence of each alias with its
associated original data type.

\myparagraph{\opDataTypeSimplification}
In programs with complex data types, such as structures, arrays, and
pointers, not all components are essential for maintaining bug-triggering
properties. For example, a bug-triggering program containing a
\mycode{struct} with three integer members can be simplified into three distinct
integer variables, and possibly only one variable is essential. To facilitate this
simplification, we introduce \opDataTypeSimplification, a strategy designed to
transform variables of complex data types into variables of primitive data
types, like integers or floats.

\myparagraph{\opVariableElimination}
Intermediate variables are pervasive in programs, and reducing them is
desirable in program reduction tasks. Besides, some variables, although not
being used, are hard to eliminate. For
instance, to remove an unused parameter, both the parameter defined in
the function and its corresponding argument passed to the call site of this
function should be removed simultaneously. This is hard or even impossible
for language-generic reduction tools. Therefore, we propose
\opVariableElimination to optimize out both intermediate and unused
variables.

The proposed transformations are universally applicable across various
programming languages, offering a broad utility.
By performing these
transformations on programs via
the \llm, substantial human effort is saved from designing and implementing
reducers that target these transformations.
While certain existing language-specific reducers like \creduce, may
already incorporate some of these transformations, \eg, \opFunctioninlining and
\opVariableElimination,
creating new transformation
passes remains a non-trivial task for users. Our approach not only
simplifies this process but also extends its reach across multiple
programming languages.

\subsection{Multi-level Prompts}
\label{subsection:prompts}
Prompts enable \llms to apply the transformations mentioned above.
We take \opFunctioninlining as an example.
We avoid directly instructing the \llm to perform transformations
exhaustively, such as inlining all functions in a program in a single query,
which might overwhelm its processing capabilities, especially for programs with
multiple functions. Instead, we employ a multi-level prompting approach.
\cref{fig:prompts} presents an example of \opFunctioninlining.
First,
we pose a primary question to the \llm (step \circled{1}): ``Given the
following
program \{ PROGRAM \}, identify all functions that can be inlined.''
Based on the list
provided by the \llm (step \circled{2}), we then ask a series of
follow-up questions (step \circled{3} and step \circled{5}) like ``Given the
following program \{ PROGRAM \}
and the specified function \{ fn1 \}, optimize \{ fn1 \} out via function
inlining.'', and the \llm do the transformations accordingly (step \circled{4}
and
step \circled{6}). This
strategy excludes irrelevant context and ensures that the queries are more
targeted, thereby increasing
the likelihood of the \llm generating high-quality results.
For other transformations, the prompts follow a similar template --- first
prompting the \llm to identify a target list, and
then instructing it to attempt optimization of each target.

%% file: figures/code_llvm_31259_loop_unrolling.tex
\begin{figure*}[ht]
	\centering
\begin{subfigure}[b]{0.3\linewidth}
        \begin{lstlisting}[language=C,  escapechar=@, numbers=left,
            numbersep=1em, commentstyle=\color{darkgray},
            xleftmargin=1em, tabsize=1]
......
// nested loop
for (i = 0; i < 7; i++)
  for (j = 0; j < 5; j++)
    for (k = 0; k < 7; k++)
      fn8(ad[i][j][k], "g_643[i][j][k]", aj);
......
\end{lstlisting}
\caption{Original}
\label{subfig:loop_unrolling:original}
\hrulefill
\end{subfigure}
    \hfil
    \verticalline
\begin{subfigure}[b]{0.3\linewidth}
        \begin{lstlisting}[language=C,  escapechar=@, numbers=left,
            numbersep=1em, commentstyle=\color{darkgray},
            xleftmargin=1em, tabsize=1]
......
// nested loop
for (i = 0; i < 7; i++)
  for (j = 0; j < 5; j++)
    for (k = 0; k < 7; k++)
      s = s ^ ad[i][j][k];
......
\end{lstlisting}
\caption{\proj: Before \opLoopUnrolling}
\label{subfig:loop_unrolling:before_op}
\hrulefill
\end{subfigure}
    \hfil
    \verticalline
\begin{subfigure}[b]{0.3\linewidth}
        \begin{lstlisting}[language=C,  escapechar=@, numbers=left,
            numbersep=1em, commentstyle=\color{darkgray},
            xleftmargin=1em, tabsize=1]
......
// the nested loop is fully unrolled
// into hundreds of lines
s = s ^ ad[2][0][5];
s = s ^ ad[2][0][6];
s = s ^ ad[2][1][0];
s = s ^ ad[2][1][1];
......
\end{lstlisting}
\caption{\proj: After \opLoopUnrolling}
\label{subfig:loop_unrolling:after_op}
\hrulefill
\end{subfigure}

\begin{subfigure}[b]{0.3\linewidth}
        \begin{lstlisting}[language=C,  escapechar=@, numbers=left,
            numbersep=1em, commentstyle=\color{darkgray},
            xleftmargin=1em, tabsize=1]
......
// all lines except for the bug-triggering one
// is removed by Perses
s = ad[2][1][0];
......
\end{lstlisting}
\caption{Final result of \proj}
\label{subfig:loop_unrolling:after_perses}
\end{subfigure}
    \hfil
    \verticalline
\begin{subfigure}[b]{0.3\linewidth}
        \begin{lstlisting}[language=C,  escapechar=@, numbers=left,
            numbersep=1em, commentstyle=\color{darkgray},
            xleftmargin=1em, tabsize=1]
......
for (i = 0; i < 7; i++)
  for (j = 0; j < 5; j++)
    for (k = 0; k < 7; k++)
      fn8(ad[i][j][k], "g_643[i][j][k]", aj);
......
\end{lstlisting}
\caption{Final result of \vulcan}
\label{subfig:loop_unrolling:vulcan}
\end{subfigure}
    \hfil
    \verticalline
\begin{subfigure}[b]{0.3\linewidth}
        \begin{lstlisting}[language=C,  escapechar=@, numbers=left,
            numbersep=1em, commentstyle=\color{darkgray},
            xleftmargin=1em, tabsize=1]
......
for (; h < 7; h++) {
  j = 0;
  for (; j < 5; j++)
    printf("%
}
......
\end{lstlisting}
\caption{Final result of \creduce}
\label{subfig:loop_unrolling:creduce}
\end{subfigure}

\caption{
	Code snippet from \llvm-31259, showcasing the original code, the effectiveness
	of \opLoopUnrolling, and the final results by \proj, \vulcan
	and \creduce.
}
\label{fig:code_llvm_31259_loop_unrolling}

\end{figure*}

%% file: plots/workflow.tex
\begin{figure}[h]
    \centering
    \includegraphics[width=\linewidth]{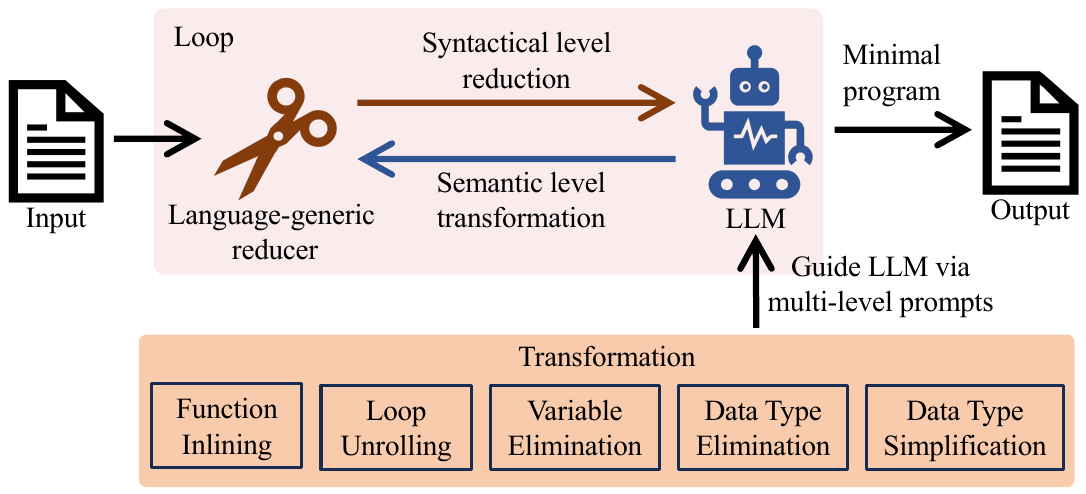}
    \caption{The workflow of \proj.
    }
    \label{fig:workflow}
\end{figure}

%% file: alg/workflow.tex
\begin{algorithm}[h]
    \small
	\DontPrintSemicolon
	\SetKwInput{KwData}{Input}
	\SetKwInput{KwResult}{Output}
	\caption{\protect\AlgMain{\program, \property, \promptMath}}
	\label{alg:main}

	\KwData{\program: the program to be reduced.}
	\KwData{$\psi: \searchspace \rightarrow \boolspace$:
		the property to be preserved by \program.}
	\KwResult{\programmin:
		the reduced program that preserves the property.}

    \program $\gets$ \AlgPerses(\program, \property) \tcp*{Quickly minimize \program for \llm to process.}
	\Repeat(\tcc*[h]{Monotonically minimize the size of \program.})
	{$|\program| \ge |\programmin|$}{
		$\programmin \gets \program$ \;
		\transformationListMath $\gets$ \AlgGetTransformationList(\promptMath) \;
		\label{alg:main:get_transformation_list}
		\tcp*[l]{Iterate through each transformation.}
		\ForEach{\transformationMath $\in$ \transformationListMath}{
			\label{alg:main:loop_transformation_start}
			\primaryQuestionMath $\gets$
			\AlgGetPrimaryQuestion(\transformationMath) \;
			\label{alg:main:get_primary_question}
			\followupQuestionMath $\gets$
			\AlgGetFollowupQuestion(\transformationMath) \;
			\label{alg:main:get_followup_question}
			\tcp*[l]{Ask \llm to identify a list of targets.}
			\label{alg:main:get_target_list}
			\targetListMath $\gets$ \AlgGetTargetList{\program,
			\primaryQuestionMath} \;
			\ForEach{\targetMath $\in$ \targetListMath}{
				\label{alg:main:loop_target_start}
				\tcp*[l]{Let \llm apply the transformation on the target.}
				\programtmp $\gets$ \AlgApplyTransformation(\program,
				\followupQuestionMath,
				\targetMath) \;
				\label{alg:main:perform_transformation}
				\If{\property(\programtmp)}{
					\program $\gets$ \programtmp \tcp*{\programtmp{} preserves the property.}
				}
			\label{alg:main:loop_target_end}
			}
			\tcp*[l]{Run \perses for further reduction.}
			\label{alg:main:perses}
			\program $\gets$ \AlgPerses(\program, \property) \;
		}
		\label{alg:main:loop_transformation_end}
	}

	\Return \programmin \;
\end{algorithm}

%% file: plots/prompts.tex
\begin{figure*}[h]
    \centering
    \includegraphics[width=0.9\textwidth]{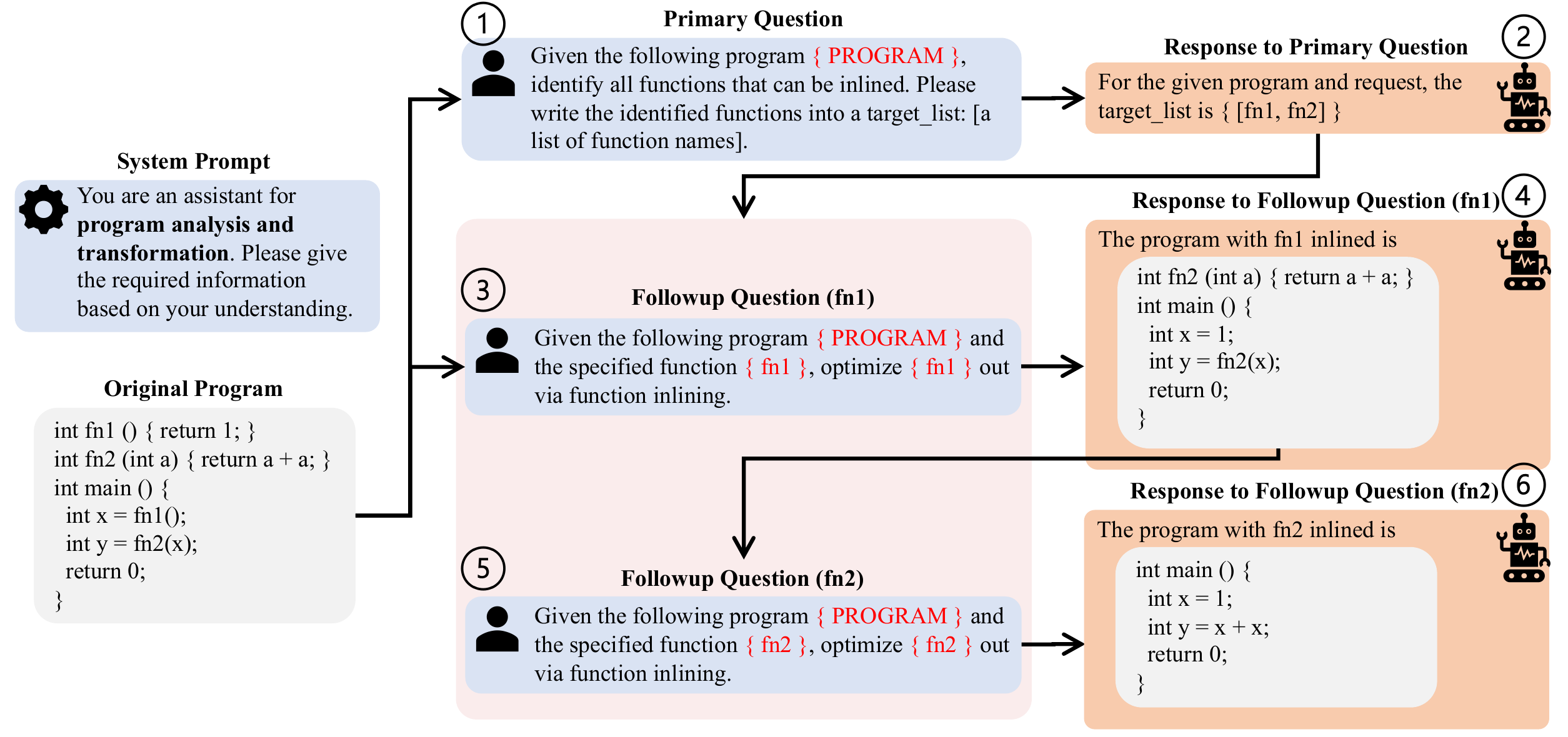}
    \caption{An example of prompt design.
    \includegraphics[width=0.4cm]{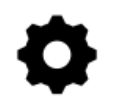} and
    \includegraphics[width=0.4cm]{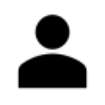} denote system prompt
    and user prompt provided by the users.
    \includegraphics[width=0.4cm]{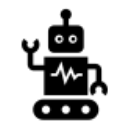} denotes the responses
    from the \llm. }
    \label{fig:prompts}
\end{figure*}

%% file: evaluation.tex
In this section, we evaluate the reduction effectiveness and efficiency of
\proj.
Specifically, we conducted the following research questions.
\begin{description}
    \item [RQ1] What is the effectiveness of \proj in program reduction?
    \item [RQ2] To what extent is the efficiency of \proj in program reduction perceived by users?
    \item [RQ3] What is the effectiveness of each transformation in \proj?
\end{description}

\subsection{Experimental Setup}
Within the workflow of \proj, we employ \perses~\cite{sun2018perses} as the
language-agnostic reducer due to its superior efficiency compared to \vulcan.
Additionally, we utilize \openaiapi~\cite{openaiapi}, specifically the
\gptCurrentVersion version, to serve as the \llm.
We also develop a variant named \projPlusVulcan,
which invokes \vulcan to further reduce the program after \proj finishes.
The experiments are conducted on
an Ubuntu 22 server
with an Intel(R) Xeon(R) Gold
6348 CPU @ 2.60GHz and 512 GB RAM.
For a fair comparison,
all algorithms were executed in a single-process, single-thread
environment.

\myparagraph{Benchmarks}
To measure the effectiveness and efficiency of \proj across various
languages, we employ three benchmark suites: \benchmarkC,
\benchmarkRust and
\benchmarkJs. The \benchmarkC, previously collected and utilized by
previous studies~\cite{xu2023pushing,rcc,zhang2023ppr}, comprises 20
large complex programs triggering real-world bugs in \llvm or \gcc.
\benchmarkRust, incorporating \benchmarkSizeRust bug-triggering \rust programs, has also
been used in prior research~\cite{xu2023pushing}.
We further craft \benchmarkJs, a non-public benchmark suite, for this
study.
Specifically,
we use FuzzJIT~\cite{wang2023fuzzjit}
to fuzz
a prevalent \javascript engine, \ie, \javascriptcore
(version c6a5bcc), and then
randomly collect 10 programs that cause miscompilations in JIT
compiler.
Since the programs and reduced programs in \benchmarkJs
are not publicly
accessible at the time of \llms' training,
and thus not in the training sets of \llms.
The evaluation on \benchmarkJs
helps to investigate whether
\proj suffers from the data leakage
problem~\cite{wu2023effective}.
In total, the evaluation benchmarks encompass \benchmarkSizeAll
programs triggering real-world bugs in compilers, spanning across three
popular programming languages.

\myparagraph{Baselines}
In all three benchmark suites, we use \perses and \vulcan as baselines.
\perses stands out as a highly effective and efficient program reduction
tool. To avoid the occurrence of syntactical invalid variants during the
reduction process, it transforms and normalizes the formal syntax of a
programming language. \vulcan~\cite{xu2023pushing}, building upon
\perses,
provides three manually designed auxiliary reducers to further
search for reduction opportunities on results from \perses.
Compared to \perses, \vulcan achieves a reduction in the number of tokens,
albeit at the expense of increased running time.
They are both language-generic and are applicable across a broad
spectrum of programming languages. We also include
\creduce (v2.9.0) as an additional baseline. \creduce not only stands as the most
effective algorithm for \candcpp, it can also be applied to other
languages, though not customized for them.

\myparagraph{Configuration}
If not otherwise specified, our experiments are conducted by invoking \openaiapi
(version \gptCurrentVersion), with the
proposed multi-level prompt and transformations
in~\cref{subsection:transformations} and ~\cref{subsection:prompts}.
To effectively harness the inherent randomness of \llms, we set
\mycode{temperature=1.0}. This high value
encourages the \llm to generate more
diverse outcomes~\cite{openaiapiTemperature}.
Additionally, we employ
\mycode{n=5} to generate five distinct results for every query~\cite{openaiapiN},
enabling us to
choose the smallest program preserving the property \property
as the optimal result. All the rest
configurations are set to their default values.

\subsection{RQ1: Reduction Effectiveness}
We measure the effectiveness of \proj, \projPlusVulcan and baseline
algorithms via the final
program sizes in tokens. A smaller size is favored, as it signifies the removal
of more bug-irrelevant code, thereby saving developers more manual effort.
The effectiveness of each algorithm on all
three benchmark suites is presented in \cref{table:effectiveness}.
Due to the randomness of \proj, we repeat five times for \proj and
\projPlusVulcan on every benchmark,
and display the mean value and standard deviation value in the table.
On each benchmark,
the minimal results are highlighted in bold.

\input{tables/effectiveness}

\myparagraph{\benchmarkC}
On this benchmark suite, \perses reduces the programs to an average of
\cSizeAvgPerses
tokens.
Building upon this, \vulcan further compresses
the average
program size into \cSizeAvgVulcan, \ie, thereby continues to decrease the program
size by \cSizeDecRateVulcanVsPerses.
Despite \vulcan's notable reduction progress, \proj is still
capable of continuing to push the limit of \perses, and reduces the
programs in \benchmarkC into \cSizeAvgProj tokens on average across all five runs.
It cuts down the
average program size of \perses by \cSizeDecRateProjVsPerses, outperforming
\vulcan
significantly by \cSizeDecRateProjVsVulcan
(proved by a \pvalue of
\cSizePValueProjVsVulcan).
\creduce stands out by
achieving the lowest average program size, \ie,  \cSizeAvgCreduce tokens.
This performance is anticipated as
\creduce incorporates various transformation passes specifically designed for
\candcpp. Despite relying on only general transformations, \projPlusVulcan still
achieves performance comparable to \creduce, averaging \cSizeAvgProjPlusVulcan
tokens.
Moreover, it outperforms \creduce in
\cMinNumOfSizeProjPlusVulcan out of \benchmarkSizeC benchmarks,
highlighting its effectiveness with only language-generic transformations.

\myparagraph{\benchmarkRust}
On \benchmarkRust, \perses and \vulcan produce programs with
\rustSizeAvgPerses and \rustSizeAvgVulcan tokens
on average, separately.
\proj and \projPlusVulcan further shrink the
average program
size into \rustSizeAvgProj and \rustSizeAvgProjPlusVulcan tokens.
\creduce produces the second-largest programs on average, only smaller than
\perses. This is anticipated since \creduce lacks specialized
transformations for \rust.

Further analysis into these benchmarks reveals that \vulcan and transformations in
\proj are
complementary
on \benchmarkRust.
\vulcan demonstrates greater effectiveness in reducing
programs of relatively smaller size, as highlighted by the average
original size of \rustMinNumOfSizeVulcanCaseAvgSize tokens
across \rustMinNumOfSizeVulcan programs, outperforming \proj.
In contrast, the \rustMinNumOfSizeProj programs where
\proj is
proved better than \vulcan have an average of \rustMinNumOfSizeProjCaseAvgSize
tokens, indicating its proficiency in reducing relatively larger
programs.
Our speculation is that \vulcan and \proj target different reduction opportunities.
\vulcan performs identifier/sub-tree replacement and local exhaustive search. Such
reducers, while lacking in semantic analysis, find reduction opportunities in a
"brute-force" manner and are particularly effective at
uncovering less obvious
reduction opportunities.
On the other hand, \proj employs more semantic and complex
transformations,
adeptly and systematically analyzing and reducing a complicated
program step by step.
Moreover, the results of \projPlusVulcan in the last column prove
the
complementary characteristic between \proj and \vulcan, which achieve the best in
\rustMinNumOfSizeProjPlusVulcan out of the total \benchmarkSizeRust
benchmarks.

\myparagraph{\benchmarkJs}
Programs in \benchmarkJs are much
simpler and smaller than those in the previous two benchmark suites.
Therefore, even \perses alone is capable of reducing the
programs to only \jsSizeAvgPerses tokens. Following this, \vulcan,
\proj and \projPlusVulcan
achieve \jsSizeAvgVulcan, \jsSizeAvgProj and \jsSizeAvgProjPlusVulcan tokens,
further
reducing the average results
by \jsSizeDecRateVulcanVsPerses, \jsSizeDecRateProjVsPerses
and \jsSizeDecRateProjVsPerses, separately.
Similar to its performance on \rust, \creduce cannot outperform the aforementioned
algorithms on \javascript, due to its lack of employment of \javascript's semantics.
The evaluation results also serve to demonstrate that the performance
exhibited by \proj is not attributable to data leakage.
These benchmarks were collected by the authors via fuzzing,
and the optimal results remain inaccessible to the public,
thereby precluding any possibility of \llms memorizing them.

\finding{
    \proj improves \perses
by
producing \cSizeDecRateProjVsPerses, \rustSizeDecRateProjVsPerses and \jsSizeDecRateProjVsPerses
smaller programs on three benchmarks.
Moreover, \projPlusVulcan improves \vulcan,
by \cSizeDecRateProjPlusVulcanVsVulcan,
\rustSizeDecRateProjPlusVulcanVsVulcan
and \jsSizeDecRateProjPlusVulcanVsVulcan.
On C language, \projPlusVulcan
performs comparably to \creduce,
a language-specific reducer for C language.
}

\subsection{RQ2: Reduction Efficiency}
In this research question, we measure the time elapsed
when users employ each technique for program reduction.\footnote{
Please note that
the time measured in this research question
cannot be directly used to compare the
computational resources consumed by \proj
with those consumed by other techniques.
\proj is standing on the
shoulders of giants,
\ie, \llms,
of which computations heavily rely on GPUs,
and it is infeasible for us to measure the actual resource consumption
of each query to \openaiapi.
In contrast,
\perses, \vulcan, and \creduce only
leverage CPUs.}
Shorter time indicates higher efficiency, and
\cref{table:efficiency} shows the results.
Since both \vulcan
and \proj perform reduction on top of \perses's results, it is impossible for these two
algorithms to take less time than \perses. In addition, as a highly
efficient
tree-based
reduction algorithm, \perses is generally faster than \creduce. Therefore, we focus
on the comparison among \vulcan, \creduce and \proj.

\input{tables/efficiency}

In \benchmarkC, compared to \vulcan and \proj, \creduce
generally has a
shorter reduction time. This is expected, as \creduce's transformations are
specifically designed for \candcpp languages, whereas \vulcan approaches the
problem in a more unguided and brute-force manner, and \proj's attempts are not
specifically designed for \candcpp and rely more on the efficiency of \llms.
On average, \proj takes \cTimeFormatAvgProj, which is
\cTimeDecPercentageOfAvgProjVsVulcan shorter than
\cTimeFormatAvgVulcan of
\vulcan.
However, in terms of the average percentage difference of each
benchmark,
\proj requires \cTimeIncAvgOfPercentageProjVsVulcan more time compared to
\vulcan in \benchmarkC.
The main reason for such a result is that \proj is more efficient
than \vulcan when
the program is large and complex,
as the transformations it performs are aware of
the semantics and have a higher success rate in reducing the program.
However, when the program is small and simple,
\vulcan can finish quickly since its search space becomes considerably small,
but for \proj, the time consumed by the \llm is not decreased significantly
and becomes dominant, making \proj less efficient than \vulcan in
these benchmarks.
On \benchmarkRust,
The results indicate a similar trend, \ie,
\proj tends to be more efficient when reducing complex programs while
\vulcan is more efficient in small and simple benchmarks.
If we only keep
benchmarks where both tools take longer than one hour, \proj requires
\cTimeDecAvgOfPercentageOneHourProjVsVulcan and
\rustTimeDecAvgOfPercentageOneHourProjVsVulcan less time compared to
\vulcan.

In our analysis of \benchmarkRust, we observed that both \proj and \vulcan can
consume an extremely long time on certain benchmarks. For
instance, \vulcan
requires 20 hours for \rust-99830, while \proj takes a similar duration on
\rust-66851
in a specific run. This extensive time consumption is often due to the frequent
invocation of \perses, which only achieves marginal progress with each transformation.
The prolonged duration of \perses is primarily attributed to the strict syntax of
the \rust language. Considering that program reduction is an NP-complete problem,
this inefficiency might be optimized in the future, whereas it
cannot be
completely eliminated.

After further in-depth analysis, another interesting fact emerges. On the three
benchmark suites, the average time taken by \proj is \cTimeAvgProj,
\rustTimeAvgProj,
and \jsTimeAvgProj hours, respectively. However, within these durations, the time
spent
waiting for the \llm responses accounts
for \cGptTimePercentProj, \rustGptTimePercentProj,
and \jsGptTimePercentProj.
Additionally, we measure the efficiency from more perspectives.
Take \benchmarkC as an example: the average expense to finish each
benchmark amounts to \$0.42, with each benchmark requiring 41
queries on average, and every query consuming approximately
39
seconds on average.
This experiment involves
invoking the \openaiapi, and might have been
limited by high user traffic and few computational resources allocated. Considering
the ongoing advancements in \llms
technology, we believe that \proj's efficiency will be substantially improved
in the future.

\finding{
    From the perspective of users, \proj is more efficient than \vulcan on
    more complex programs with longer
    processing time, while \vulcan reduces faster than \proj on simpler and shorter
    programs.
}

\subsection{RQ3: Effectiveness of Each Transformation}
\input{plots/size_changes_boxplot}

To answer this question, we delve into the impact of each transformation on program
size, alongside their potential to escape the local minimal program and unlock new
reduction opportunities.
Our in-depth analysis focuses on \benchmarkC
because of
its complexity of bug-triggering programs
compared to other benchmark suites,
ensuring enough room for each
transformation to take effect.

For each transformation, we measure how the program size
changes in each benchmark of
\benchmarkC after a specific
transformation is performed, and plot the size changes into a box-plot and red dots, as
shown in~\cref{fig:size_changes_boxplot}. Additionally, since each transformation is
immediately
followed by an invocation of \perses, we also monitor the cumulative size changes
resulting from both the transformation and its subsequent \perses reduction. These
changes are depicted by the blue dots in the second box-plot of each subplot in
~\cref{fig:size_changes_boxplot}. Given that a transformation is performed
in every
iteration, we
compute the average size change by dividing the total sum of size changes for that
transformation by the number of iterations.
This approach enables us to thoroughly comprehend how each transformation influences
program size and assess its capacity to provide further reduction opportunities to
\perses.

According to size changes induced by each transformation alone, \ie, the left
box-plot with red dots, we can find two trends.
First, \opFunctioninlining, \opDataTypeElimination and \opVariableElimination are
more likely to reduce the program size by themselves, with an
average size
change
\cSizeChangeFIAlone, \cSizeChangeDTEAlone and \cSizeChangeVEAlone,
separately. However, for the rest two transformations, \ie, \opLoopUnrolling and
\opDataTypeSimplification, most of the program sizes increase instead, with an
average of \cSizeChangeLUAlone and \cSizeChangeDTSAlone respectively. This is
expected,
as such transformations will generally transform the program into a larger one.
\opLoopUnrolling, disassembles a loop into
repeated lines of code, and leads to size increase temporarily.
\opDataTypeSimplification can dismantle a variable in a complex
data type, \eg, structure,
into a list of members in primary data types, which may require
more tokens to
declare and initialize each variable.

For size changes induced by both a transformation and the subsequent execution of
\perses, \ie, the right
box-plot with blue dots,
all of the proposed transformation result in size decreases.
The fact that the right box-plot is generally lower than the left one
indicates that
\perses often further removes tokens after the transformation is applied.
For \opLoopUnrolling and \opDataTypeSimplification,
even though they usually
introduce more tokens to the program,
they expose reduction opportunities for the following execution of \perses,
and eventually result in a smaller program.

To further understand the impact of each transformation on the entirety of \benchmarkC,
we provide a detailed analysis of their contributions. Specifically, on the 5 repeated
experiments on \benchmarkSizeC benchmarks, we calculate the
average size reduction brought about by each transformation across all these
100 runs by
summing up the size decreases attributed to the transformation in each benchmark and
then computing the mean value, as illustrated
in~\cref{fig:size_changes_piechart}.
Additionally, we quantify the prevalence of each transformation by counting the number
of benchmarks in which it induces a size decrease, as demonstrated
in~\cref{fig:size_changes_barchart}. The above evaluations
together offer a
comprehensive view of how each transformation influences program sizes
and
how frequently they take effect within \benchmarkC.

From~\cref{fig:size_changes_piechart}, it is evident that every transformation
contributes to further size reduction in \benchmarkC. Specifically, while
\opFunctioninlining is responsible for a reduction of \cSizeDecAllFI tokens on
average, contributing \cSizeDecAllPercentFI to the overall
decrease.
\opLoopUnrolling shows a minimal effect, contributing only
\cSizeDecAllLU
tokens, \cSizeDecAllPercentLU to the overall decrease. This highlights the varying
degrees of influence each
transformation has on program size.
Further insights from \cref{fig:size_changes_barchart} reveal that
\opDataTypeElimination affects all benchmarks, likely due to the ubiquitous presence of
\mycode{typedef} across all benchmarks. In contrast, \opLoopUnrolling is the least
prevalent transformation, affecting merely \cSizeDecFrequencyLU on average out of
\benchmarkSizeC benchmarks. This outcome is expected, considering
that not all programs involve loop structures, and not every loop is
irrelevant to the compiler
bugs. In addition, the relatively higher standard deviations
observed in the
\opDataTypeSimplification and \opLoopUnrolling suggest that these
transformations present greater challenges for the \llm to
effectively handle.

\input{plots/size_changes_summary}

\finding{In \benchmarkC, all proposed transformations contribute to the further reduction
by either
shrinking the programs directly or providing reduction opportunities to \perses. }

%% file: tables/effectiveness.tex
\begin{table}[h!]
	\centering
   	\caption{The reduction sizes
            of \perses, \vulcan,
   	\creduce, \proj and \projPlusVulcan.
    Best results among all algorithms are highlighted in bold
   	font.
       }
	\label{table:effectiveness}
\resizebox{\linewidth}{!}{%
\begin{tabular}{@{}c|c|r|
                >{\columncolor[HTML]{B7B7B7}}r
                >{\columncolor[HTML]{D9D9D9}}r
                >{\columncolor[HTML]{F3F3F3}}r
                >{}r @{${}\pm{}$}
                >{}l
                >{}r @{${}\pm{}$}
                >{}l}
    \toprule
\multicolumn{1}{l|}{}                              &  Benchmark   & \multicolumn{1}{c|}{Original} &
\multicolumn{1}{c}{\cellcolor[HTML]{B7B7B7}\perses} &
\multicolumn{1}{c}{\cellcolor[HTML]{D9D9D9}\vulcan}  &
\multicolumn{1}{c}{\cellcolor[HTML]{F3F3F3}\creduce} &
\multicolumn{2}{c}{\proj}             & \multicolumn{2}{c}{\proj{}+\vulcan}          \\ \midrule
& \llvm-22382  &                        21,068 & 144                                                 &
108                                               & 70     & 73.2           &
1.6                                                                & \textbf{69.8}  & 1.8                                          \\
& \llvm-22704  &                       184,444 & 78                                                  &
62                                                & 42     & 43.6           &
3.1                                                                & \textbf{41.8}  & 3.3                                          \\
& \llvm-23309  &                        38,647 & 464                                                 &
303                                               & 118             & 105.8 &
9.3                                                                & \textbf{91.2}  & 8.2                                          \\
& \llvm-23353  &                        30,196 & 98                                                  &
91                                                & 74              & 68.8  &
8.0                                                                & \textbf{66.6}  & 6.5                                          \\
& \llvm-25900  &                        78,960 & 239                                                 &
104                                               & 90     & 93.4           &
11.1                                                               & \textbf{84}    & 6.5                                          \\
& \llvm-26760  &                       209,577 & 120                                                 & 56                                                & \textbf{43}     & 62.8           & 18.6                                                               & 52.6  & 5.4                                          \\
& \llvm-27137  &                       174,538 & 180                                                 & 88                                                & \textbf{50}     & 69.2           & 19.2                                                               & 65    & 20.2                                         \\
& \llvm-27747  &                       173,840 & 117                                                 &
79                                                & 68     & 87.8           &
2.5                                                                & \textbf{63.2}  & 2.2                                          \\
& \llvm-31259  &                        48,799 & 406                                                 &
282                                               & 168    & 184.0          &
51.2                                                               & \textbf{114.4} & 10.9                                         \\
&  \gcc-59903  &                        57,581 & 308                                                 & 198                                               & \textbf{105}    & 209.8          & 72.1                                                               & 166.4 & 64.0                                         \\
&  \gcc-60116  &                        75,224 & 443                                                 &
247                                               & 168    & 188.8          &
52.4                                                               & \textbf{127.6} & 34.8                                         \\
&  \gcc-61383  &                        32,449 & 272                                                 & 195                                               & \textbf{84}     & 113.2          & 13.6                                                               & 105.2 & 5.4                                          \\
&  \gcc-61917  &                        85,359 & 150                                                 & 103                                               & \textbf{65}     & 78.4           & 10.6                                                               & 73.4  & 5.5                                          \\
&  \gcc-64990  &                       148,931 & 239                                                 &
203                                               & \textbf{65}     & 143.4          &
58.0                                                               & 119   & 50.1                                         \\
&  \gcc-65383  &                        43,942 & 153                                                 &
84                                                & 72              & \textbf{64.2}  &
2.7                                                                & \textbf{64.2}  & 2.7                                          \\
&  \gcc-66186  &                        47,481 & 327                                                 &
226                                               & 115             & 97.8  &
17.9                                                               & \textbf{94.2}  & 12.0                                         \\
&  \gcc-66375  &                        65,488 & 440                                                 &
227                                               & \textbf{56}     & \textbf{56.0}  &
5.1                                                                & \textbf{56.0}    & 5.1                                          \\
&  \gcc-70127  &                       154,816 & 301                                                 &
230                                               & 84     & 95.0           &
3.8                                                                & \textbf{73.6}  & 3.3                                          \\
&  \gcc-70586  &                       212,259 & 426                                                 & 223                                               & \textbf{130}    & 235.4          & 30.2                                                               & 156.8 & 12.6                                         \\
&  \gcc-71626  &                         6,133 & 51                                                  &
38                                       & 46              & 44.6           &
3.4                                                                & \textbf{36.6}  & 0.9                                          \\
\cmidrule(l){2-10}
\multirow{-21}{*}{\rotatebox[origin=c]{90}{\benchmarkC}}   &     Mean
&                        94,487 & 247.8                                               &
157.4                                             & \textbf{85.7}     & 105.8          &
4.4                                                                & 86.1  & 2.9                                          \\ \midrule
& \rust-44800  &                           801 & 467                                                 &
284                                               & 473             & 124.6 &
32.4                                                               & \textbf{118.6} & 34.4                                         \\
& \rust-66851  &                           936 & 728                                                 &
713                                               & 654             & 414.2 &
273.9                                                              & \textbf{331.2} & 257.8
\\
& \rust-69039  &                           190 & 114                                                 &
101                                               & 110             & 97.2  &
8.0                                                                & \textbf{90.8}  & 10.9                                         \\
& \rust-77002  &                           347 & 263                                                 &
247                                               & 264             & \textbf{96.0}  &
27.9                                                               & \textbf{96.0}  & 27.9                                         \\
& \rust-77320  &                           173 & 40                                                  &
40                                       & 40     & 40.0  & 0.0                                                                &
\textbf{39.0}  & 0.0                                          \\
& \rust-77323  &                            81 & \textbf{13}                                                  &
\textbf{13}                                       & \textbf{13}     & \textbf{13.0}  &
0.0                                                                & \textbf{13.0}  & 0.0                                          \\
& \rust-77910  &                            63 & 34                                                  &
\textbf{21}                                       & 23              & 29.2           &
2.7                                                                & \textbf{21.0}  & 0.0                                          \\
& \rust-77919  &                           132 & 74                                                  &
62                                       & 70              & 62.2           &
14.3                                                               & \textbf{58.2}  & 7.7                                          \\
& \rust-78005  &                           182 & 102                                                 &
102                                               & \textbf{75}     & 102.0          &
0.0                                                                & 102.0 & 0.0                                          \\
& \rust-78325  &                            \textbf{65} & 29                                                  &
\textbf{26}                                       & 34              & 29.0           &
0.0                                                                & \textbf{26.0}  & 0.0                                          \\
& \rust-78651  &                           957 & 17                                                  & \textbf{9}                                        & 12              & 16.6           & 0.5                                                                & 11.0  & 2.7                                          \\
& \rust-78652  &                           263 & 56                                                  &
\textbf{49}                                       & \textbf{49}     & 53.6           &
2.2                                                                & \textbf{49.0}  & 0.0                                          \\
& \rust-78655  &                            28 & \textbf{26}                                                  &
\textbf{26}                                       & \textbf{26}     & \textbf{26.0}  &
0.0                                                                & \textbf{26.0}  & 0.0                                          \\
& \rust-78720  &                           121 & 72                                                  & 56                                                & \textbf{51}     & 58.4           & 2.1                                                                & 56.6  & 0.5                                          \\
& \rust-91725  &                           513 & 174                                                 &
86                                                & 101             & 68.6  &
57.6                                                               & \textbf{55.2}  & 18.4                                         \\
& \rust-99830  &                           448 & 299                                                 & 277                                               & \textbf{160}    & 271.6          & 12.5                                                               & 230.2 & 62.6                                         \\
& \rust-111502 &                           192 & 166                                                 &
157                                               & 161             & 104.8 &
7.2                                                                & \textbf{103.6} & 8.2                                          \\
& \rust-112061 &                           556 & 458                                                 &
442                                               & 450             & 385.0 &
32.6                                                               & \textbf{380.4} & 31.8                                         \\
& \rust-112213 &                           866 & 736                                                 &
635                                      & 732             & 653.0          &
34.0                                                               & \textbf{618.8} & 22.9                                         \\
& \rust-112526 &                           644 & 382                                                 &
338                                               & 545             & 304.0 &
30.3                                                               & \textbf{283.2} & 23.3                                         \\
\cmidrule(l){2-10}
\multirow{-21}{*}{\rotatebox[origin=c]{90}{\benchmarkRust}} &     Mean
&                           378 & 212.5                                               & 184.2
& 202.2           & 147.5 & 11.1                                                               & \textbf{135.5} &
10.1                                         \\ \midrule
&    \js-1     &                           244 & 52                                                  &
41                                                & 41              & 25.6  &
0.5                                                                & \textbf{24.8}  & 1.6                                          \\
&    \js-2     &                           112 & 51                                                  &
41                                                & 40              & 34.0  &
6.5                                                                & \textbf{27.8}  & 3.5                                          \\
&    \js-3     &                           125 & 57                                                  &
41                                       & 47              & 51.6           &
12.1                                                               & \textbf{35.4}  & 7.7                                          \\
&    \js-4     &                           185 & 65                                                  &
35                                                & 47              & 33.4  &
2.2                                                                & \textbf{28.2}  & 6.9                                          \\
&    \js-5     &                           178 & 66                                                  &
38                                       & 52              & 41.6           &
2.2                                                                & \textbf{33.8}  & 5.5                                          \\
&    \js-6     &                           152 & 57                                                  &
30                                                & 45              & 20.0  &
2.8                                                                & \textbf{19.2}  & 2.7                                          \\
&    \js-7     &                           144 & 46                                                  &
38                                                & 38              & 34.0  &
0.0                                                                & \textbf{27.2}  & 6.3                                          \\
&    \js-8     &                           121 & 55                                                  &
47                                                & 45              & 40.6  &
6.3                                                                & \textbf{33.0}  & 17.4                                         \\
&    \js-9     &                            87 & 50                                                  &
30                                                & 32              & 23.8  &
2.5                                                                & \textbf{18.8}  & 5.2                                          \\
&    \js-10    &                            63 & 56                                                  &
41                                                & 43              & 34.8  &
5.1                                                                & \textbf{27.0}  & 6.3                                          \\
\cmidrule(l){2-10}
\multirow{-11}{*}{\rotatebox[origin=c]{90}{\benchmarkJs}}  &     Mean
&                           141 & 55.5                                                & 38.2
& 43.0            & 33.9  & 1.6                                                                & \textbf{27.5}  &
5.9                                          \\ \bottomrule
\end{tabular}
}
\end{table}

%% file: tables/efficiency.tex
\begin{table}[h!]
	\centering
	\caption{The reduction time (in the format of hh:mm:ss) of \perses, \vulcan,
	\creduce, \proj and \projPlusVulcan.
    }
	\label{table:efficiency}
\resizebox{\linewidth}{!}{%
\begin{tabular}{@{}c|c|
                >{\columncolor[HTML]{B7B7B7}}r
>{\columncolor[HTML]{D9D9D9}}r
>{\columncolor[HTML]{F3F3F3}}r
>{}r @{${}\pm{}$}
>{}l
>{}r @{${}\pm{}$}
>{}l}
    \toprule
                                \multicolumn{1}{l|}{}                              &  Benchmark   &
                                \multicolumn{1}{c}{\cellcolor[HTML]{B7B7B7}\perses} &
                                \multicolumn{1}{c}{\cellcolor[HTML]{D9D9D9}\vulcan} &
                                \multicolumn{1}{c}{\cellcolor[HTML]{F3F3F3}\creduce} &
                                \multicolumn{2}{c}{\proj}           &
                                \multicolumn{2}{c}{\projPlusVulcan}             \\ \midrule
                                                                                   & \llvm-22382  &
                                                                                   0:06:46                                             &
                                                                                   0:17:03                                  & 0:14:46   &
                                                                                   0:39:43                                      &
                                                                                   0:17:13                                              &
                                                                                   0:44:33                                      & 0:16:46   \\
& \llvm-22704  & 0:33:58                                             & 0:38:03                                  & 0:22:38   & 0:48:53                                      & 0:01:03                                              & 0:52:17                                      & 0:01:23   \\
& \llvm-23309  & 0:22:34                                             & 2:02:39                                           & 0:48:47   & 1:35:13                             & 0:13:36                                              & 2:05:40                                      & 0:14:16   \\
& \llvm-23353  & 0:10:31                                             & 0:15:05                                  & 0:13:36   & 0:25:23                                      & 0:06:05                                              & 0:30:47                                      & 0:06:33   \\
& \llvm-25900  & 0:09:53                                             & 0:23:08                                  & 0:22:04   & 0:44:03                                      & 0:07:56                                              & 0:54:33                                      & 0:07:21   \\
& \llvm-26760  & 0:21:54                                             & 0:34:23                                  & 0:32:25   & 0:54:40                                      & 0:18:08                                              & 1:04:28                                      & 0:16:51   \\
& \llvm-27137  & 1:54:41                                             & 3:15:20                                           & 2:21:43   & 2:33:15                             & 0:09:08                                              & 3:13:24                                      & 0:08:48   \\
& \llvm-27747  & 0:13:32                                             & 0:28:02                                  & 0:25:30   & 0:32:04                                      & 0:02:28                                              & 0:45:53                                      & 0:03:17   \\
& \llvm-31259  & 0:32:30                                             & 4:03:54                                  & 1:13:20   & 4:08:08                                      & 0:34:26                                              & 5:01:39                                      & 0:54:49   \\
&  \gcc-59903  & 0:48:47                                             & 1:21:15                                  & 1:23:10   & 2:14:57                                      & 0:45:14                                              & 2:44:43                                      & 0:48:09   \\
&  \gcc-60116  & 0:36:18                                             & 2:02:01                                  & 1:15:45   & 3:16:38                                      & 0:31:53                                              & 4:25:39                                      & 0:34:36   \\
&  \gcc-61383  & 0:44:59                                             & 3:50:40                                           & 1:00:39   & 2:52:39                             & 0:30:41                                              & 4:36:47                                      & 0:22:35   \\
&  \gcc-61917  & 0:14:57                                             & 0:24:16                                  & 0:44:26   & 0:37:28                                      & 0:07:37                                              & 0:43:30                                      & 0:07:31   \\
&  \gcc-64990  & 0:51:05                                             & 1:27:12                                  & 1:20:05   & 2:03:05                                      & 0:31:38                                              & 2:22:37                                      & 0:22:33   \\
&  \gcc-65383  & 0:17:05                                             & 0:37:02                                  & 0:33:45   & 0:46:53                                      & 0:05:25                                              & 0:59:14                                      & 0:05:08   \\
&  \gcc-66186  & 0:41:19                                             & 5:35:16                                           & 1:24:13   & 2:39:23                             & 0:19:26                                              & 4:05:34                                      & 0:36:49   \\
&  \gcc-66375  & 0:46:28                                             & 3:59:57                                           & 2:02:40   & 2:30:08                             & 0:10:24                                              & 3:06:41                                      & 0:10:44   \\
&  \gcc-70127  & 0:44:47                                             & 4:45:15                                           & 1:40:01   & 2:23:13                             & 0:20:36                                              & 3:24:03                                      & 0:20:20   \\
&  \gcc-70586  & 1:33:35                                             & 6:53:31                                           & 1:37:16   & 6:24:35                             & 1:26:57                                              & 10:36:26                                     & 2:32:53   \\
&  \gcc-71626  & 0:00:40                                             & 0:01:18                                  & 0:04:06   & 0:07:38                                      & 0:01:27                                              & 0:08:11                                      & 0:01:32   \\
\cmidrule(l){2-9}
\multirow{-21}{*}{\rotatebox[origin=c]{90}{\benchmarkC}}   &     Mean     & 0:35:19                                             & 2:08:46                                            & 0:59:03  & 1:54:54                             & 0:05:17                                              & 2:37:20                                      & 0:08:05   \\ \midrule
& \rust-44800  & 0:13:32                                             & 1:58:31                                           & 1:33:17   & 1:47:29                             & 0:31:37                                              & 2:15:39                                      & 0:42:02   \\
& \rust-66851  & 0:59:47                                             & 8:49:11                                  & 1:32:02   & 11:21:39                                     & 9:56:51                                              & 16:43:40                                     & 12:54:53  \\
& \rust-69039  & 0:07:54                                             & 1:25:33                                           & 0:10:05   & 0:24:13                             & 0:05:33                                              & 0:39:22                                      & 0:05:51   \\
& \rust-77002  & 0:04:12                                             & 0:20:17                                  & 0:29:18   & 0:52:27                                      & 0:15:54                                              & 0:58:21                                      & 0:15:18   \\
& \rust-77320  & 0:00:06                                             & 0:01:22                                  & 0:01:51   & 0:02:36                                      & 0:00:32                                              & 0:04:05                                      & 0:00:32   \\
& \rust-77323  & 0:00:01                                             & 0:00:11                                  & 0:00:37   & 0:00:16                                      & 0:00:03                                              & 0:00:27                                      & 0:00:04   \\
& \rust-77910  & 0:00:08                                             & 0:00:47                                  & 0:01:12   & 0:04:58                                      & 0:01:34                                              & 0:05:44                                      & 0:01:35   \\
& \rust-77919  & 0:00:17                                             & 0:02:46                                  & 0:05:29   & 0:08:45                                      & 0:05:08                                              & 0:11:18                                      & 0:04:52   \\
& \rust-78005  & 0:00:10                                             & 0:01:57                                  & 0:02:30   & 0:10:44                                      & 0:01:32                                              & 0:12:55                                      & 0:01:31   \\
& \rust-78325  & 0:00:02                                             & 0:00:28                                  & 0:01:32   & 0:00:46                                      & 0:00:35                                              & 0:01:19                                      & 0:00:34   \\
& \rust-78651  & 0:00:04                                             & 0:00:23                                  & 0:01:09   & 0:01:33                                      & 0:00:33                                              & 0:02:02                                      & 0:00:36   \\
& \rust-78652  & 0:00:08                                             & 0:01:32                                  & 0:03:01   & 0:02:53                                      & 0:02:02                                              & 0:04:38                                      & 0:01:59   \\
& \rust-78655  & 0:00:01                                             & 0:00:49                                  & 0:01:30   & 0:02:20                                      & 0:00:31                                              & 0:03:14                                      & 0:00:32   \\
& \rust-78720  & 0:00:16                                             & 0:03:47                                  & 0:06:31   & 0:11:44                                      & 0:04:47                                              & 0:13:52                                      & 0:04:55   \\
& \rust-91725  & 0:03:36                                             & 0:17:48                                           & 0:37:19   & 0:16:13                             & 0:03:55                                              & 0:23:29                                      & 0:02:11   \\
& \rust-99830  & 0:48:23                                             & 20:08:32                                          & 11:12:22  & 4:23:44                             & 1:02:03                                              & 24:28:10                                     & 1:37:03   \\
& \rust-111502 & 0:00:55                                             & 0:10:35                                  & 0:10:23   & 0:37:29                                      & 0:11:08                                              & 0:44:39                                      & 0:11:13   \\
& \rust-112061 & 0:34:50                                             & 4:48:04                                  & 1:15:44   & 5:10:02                                      & 3:12:51                                              & 8:01:03                                      & 3:04:33   \\
& \rust-112213 & 0:56:40                                             & 15:21:26                                          & 1:08:21   & 7:33:21                             & 2:44:52                                              & 17:07:36                                     & 3:34:22   \\
& \rust-112526 & 0:45:36                                             & 2:55:07                                  & 1:35:52   & 3:33:13                                      & 1:21:21                                              & 4:52:26                                      & 1:54:13   \\
\cmidrule(l){2-9}
\multirow{-21}{*}{\rotatebox[origin=c]{90}{\benchmarkRust}} &     Mean     & 0:13:50                                             & 2:49:27                                           & 1:00:30   & 1:50:19                             & 0:40:46                                              & 3:51:42                                      & 0:44:16   \\ \midrule
&    \js-1     & 0:01:29                                             & 0:29:19                                           & 0:06:59   & 0:11:47                             & 0:05:01                                              & 0:14:32                                      & 0:05:00   \\
&    \js-2     & 0:02:15                                             & 0:17:10                                           & 0:11:47   & 0:14:36                             & 0:01:12                                              & 0:29:51                                      & 0:04:21   \\
&    \js-3     & 0:01:29                                             & 0:26:35                                           & 0:06:02   & 0:09:00                             & 0:00:14                                              & 0:29:30                                      & 0:07:58   \\
&    \js-4     & 0:00:19                                             & 0:15:43                                           & 0:01:44   & 0:07:22                             & 0:02:35                                              & 0:37:07                                      & 0:03:57   \\
&    \js-5     & 0:00:14                                             & 0:04:25                                  & 0:01:39   & 0:10:32                                      & 0:04:27                                              & 0:22:49                                      & 0:04:32   \\
&    \js-6     & 0:02:59                                             & 0:23:20                                           & 0:11:33   & 0:13:24                             & 0:07:49                                              & 0:18:43                                      & 0:08:08   \\
&    \js-7     & 0:00:15                                             & 0:03:30                                  & 0:16:21   & 0:05:28                                      & 0:01:06                                              & 0:10:52                                      & 0:01:09   \\
&    \js-8     & 0:00:44                                             & 0:13:48                                           & 0:02:21   & 0:08:18                             & 0:02:20                                              & 0:13:04                                      & 0:02:10   \\
&    \js-9     & 0:01:32                                             & 0:14:14                                           & 0:07:50   & 0:09:09                             & 0:02:43                                              & 0:15:14                                      & 0:03:40   \\
&    \js-10    & 0:01:31                                             & 0:17:44                                           & 0:07:24   & 0:15:04                             & 0:04:13                                              & 0:32:32                                      & 0:05:37   \\
\cmidrule(l){2-9}
\multirow{-11}{*}{\rotatebox[origin=c]{90}{\benchmarkJs}}  &     Mean     & 0:01:17                                             & 0:16:35                                             & 0:07:22                             & 0:10:28 & 0:01:26                                              & 0:22:25                                      & 0:02:33    \\ \bottomrule
\end{tabular}
}
\end{table}

%% file: plots/size_changes_boxplot.tex
\begin{figure*}[htbp]
	\centering
	\begin{subfigure}[t]{0.17\linewidth}
		\centering
		\includegraphics[width=\linewidth, height=3.05cm]{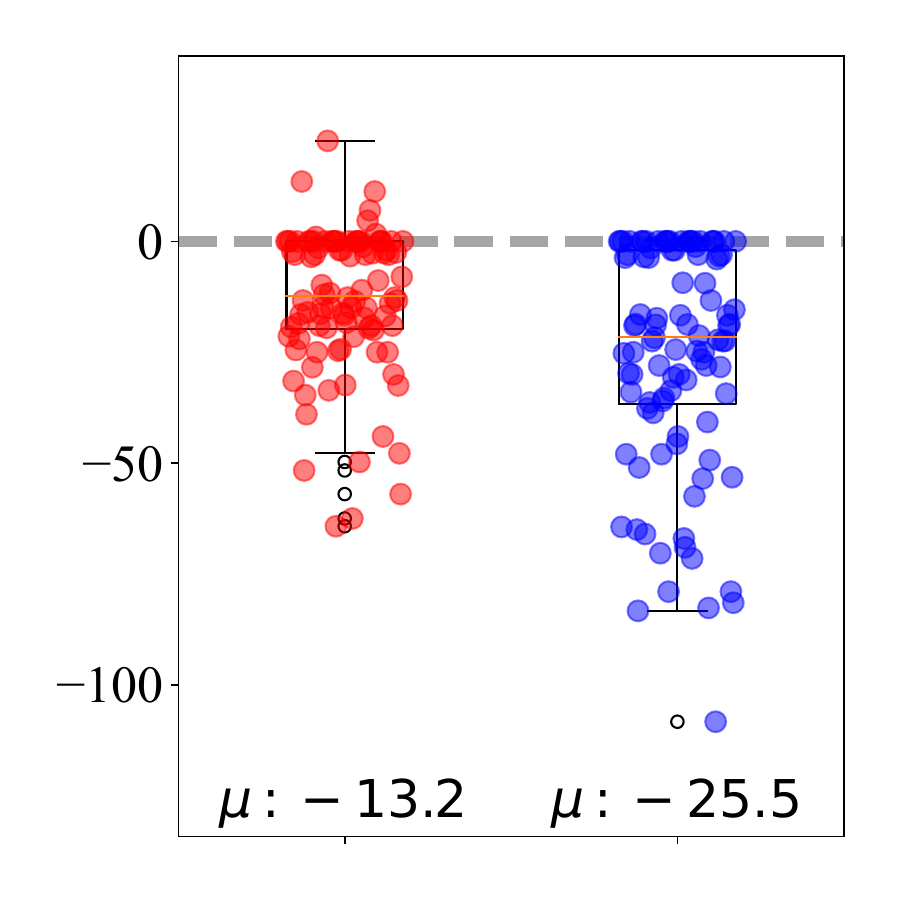}
		\caption{\opFunctioninlining}
		\label{fig:size_change_function_inlining}
	\end{subfigure}
    \hfil
	\begin{subfigure}[t]{0.17\linewidth}
		\centering
		\includegraphics[width=\linewidth,height=3.05cm]{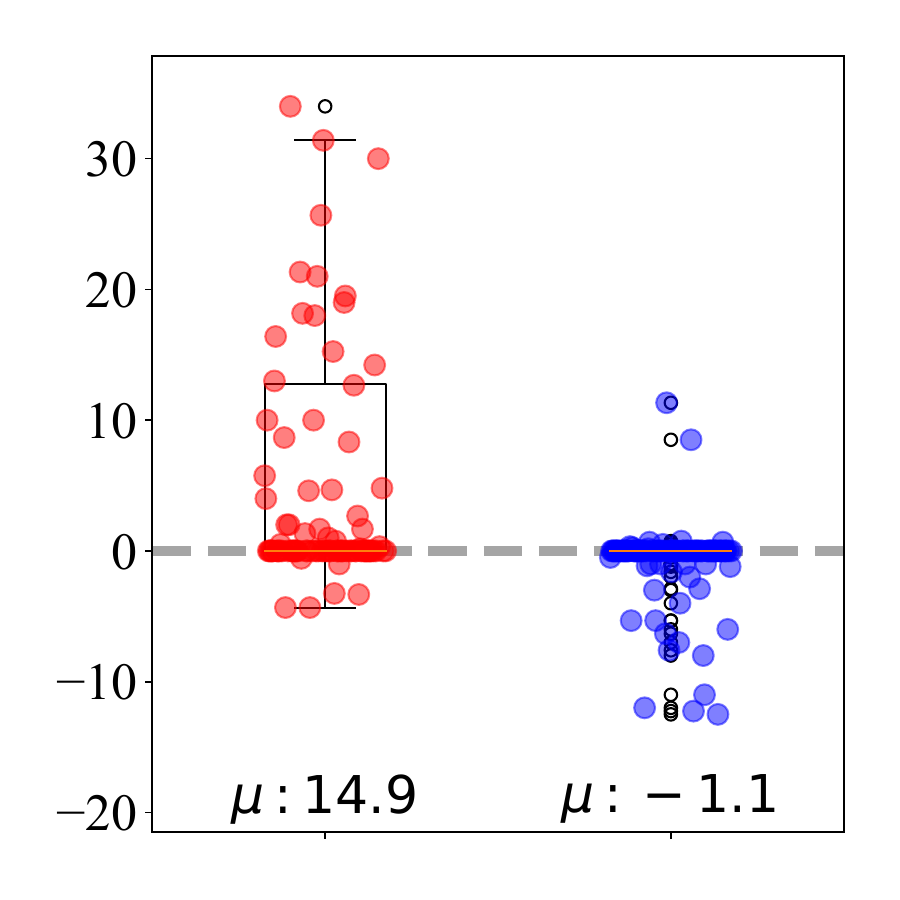}
		\caption{\opLoopUnrolling}
		\label{fig:size_change_loop_unrolling}
	\end{subfigure}
    \hfil
		\begin{subfigure}[t]{0.21\linewidth}
		\centering
		\includegraphics[width=0.809\linewidth, height=3.05cm]{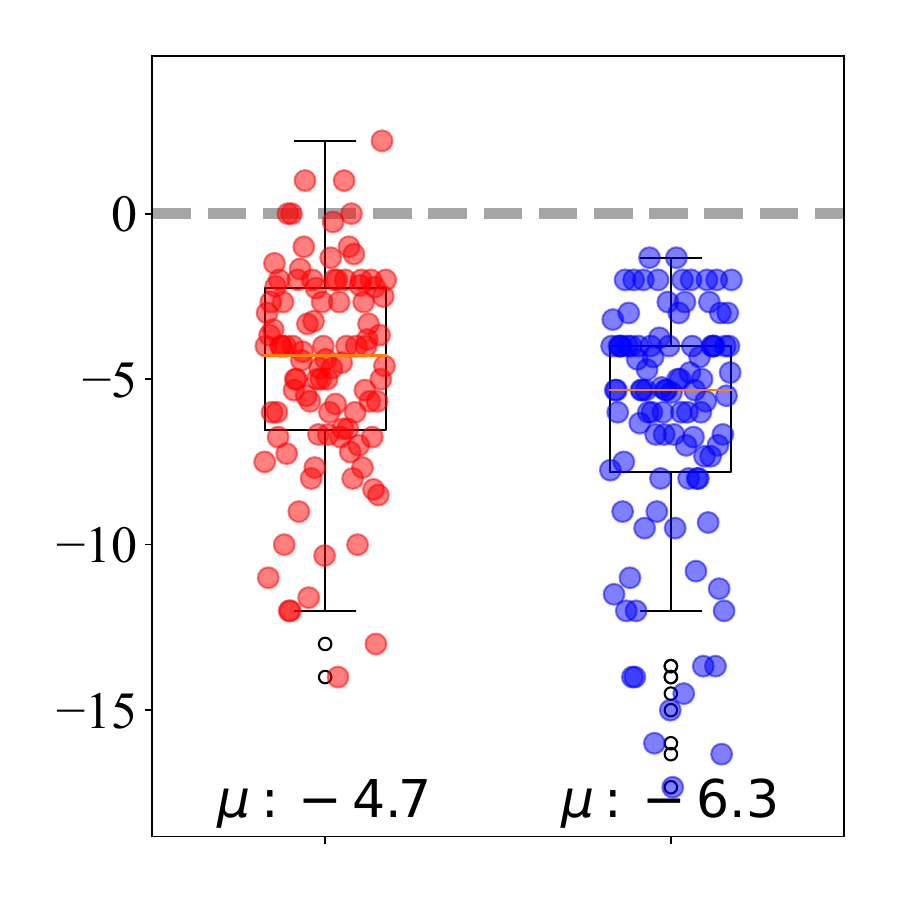}
		\caption{\opDataTypeElimination}
		\label{fig:size_change_redundant_data_type}
	\end{subfigure}
	\hfil
	\begin{subfigure}[t]{0.21\linewidth}
		\centering
		\includegraphics[width=0.809\linewidth, height=3.05cm]{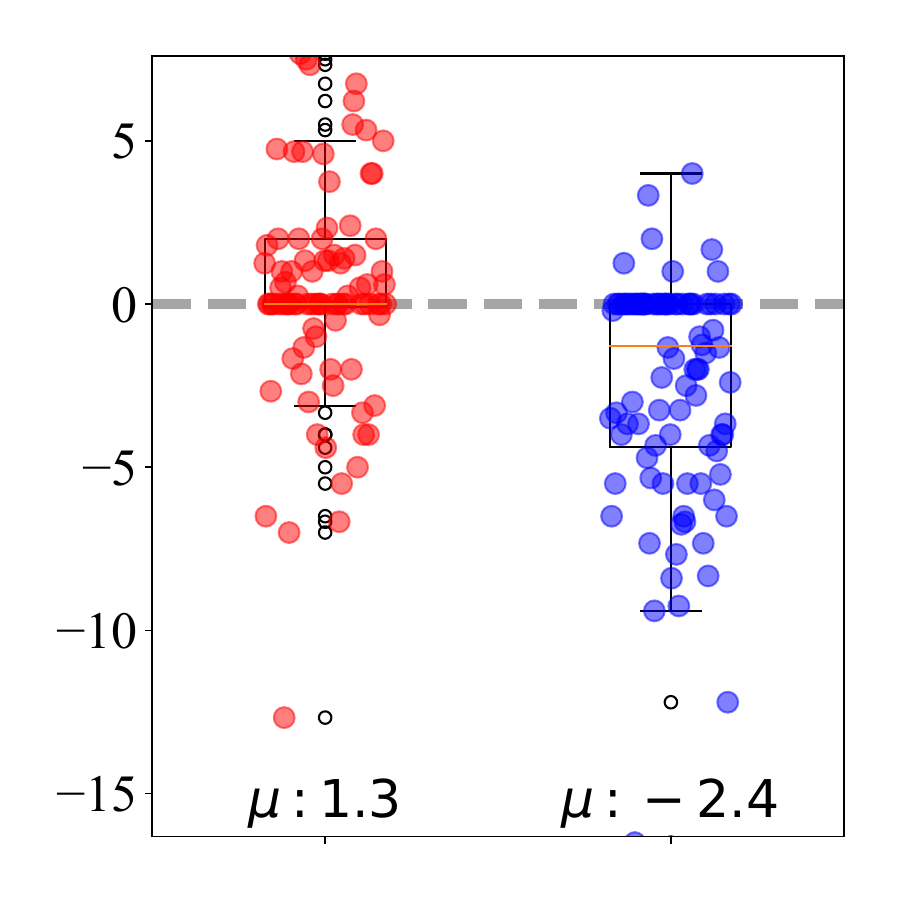}
		\caption{\opDataTypeSimplification}
		\label{fig:size_change_complex_data_type}
	\end{subfigure}
    \hfil
	\begin{subfigure}[t]{0.17\linewidth}
		\centering
		\includegraphics[width=\linewidth,height=3.05cm]{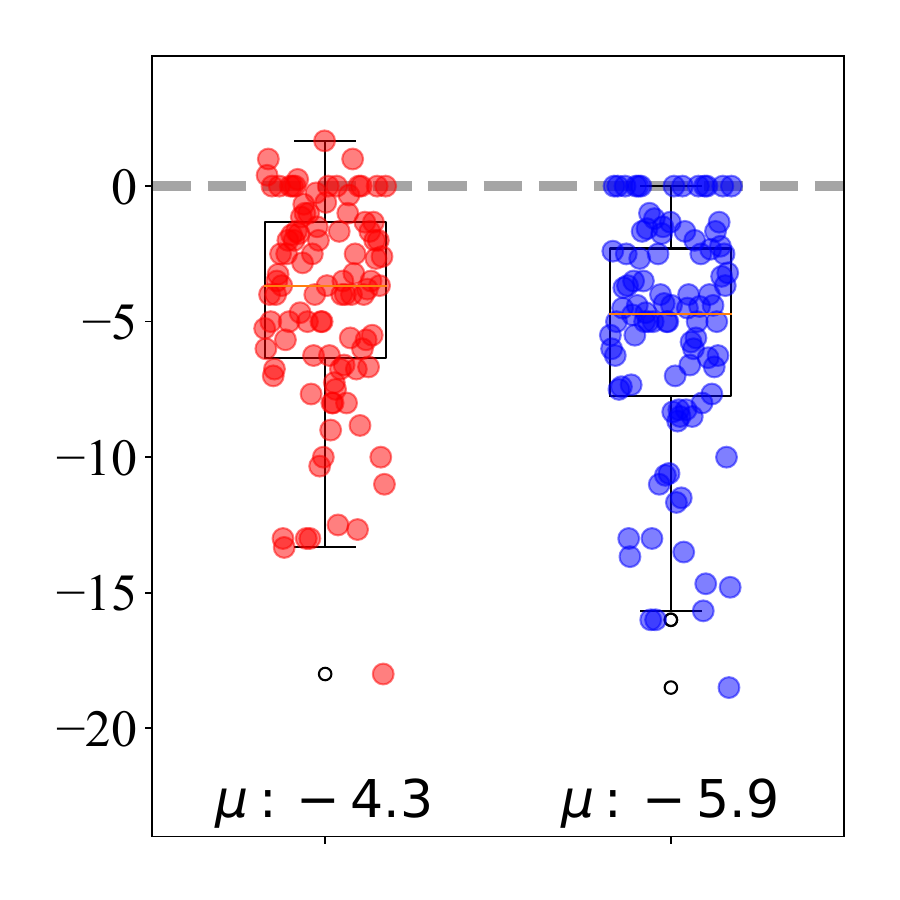}
		\caption{\opVariableElimination}
		\label{fig:size_change_unnecessary_variable}
	\end{subfigure}
	\caption{
		Program size changes induced by each transformation on benchmarks of \benchmarkC.
		In each subplot, the left box-plot and red dots represent how the size of each
		program changes before and after executing the transformation. The right
		box-plot and blue dots represent the size change of each benchmark after
		executing the transformation and the follow-up \perses reduction. There are a total of
		\benchmarkSizeC benchmarks in \benchmarkC, and each experiment is
		repeated 5 times. Therefore, we draw 100 data points on each boxplot.
	}
	\label{fig:size_changes_boxplot}
\end{figure*}

%% file: plots/size_changes_summary.tex
\begin{figure}
        \centering
        \includegraphics[width=0.9\linewidth]{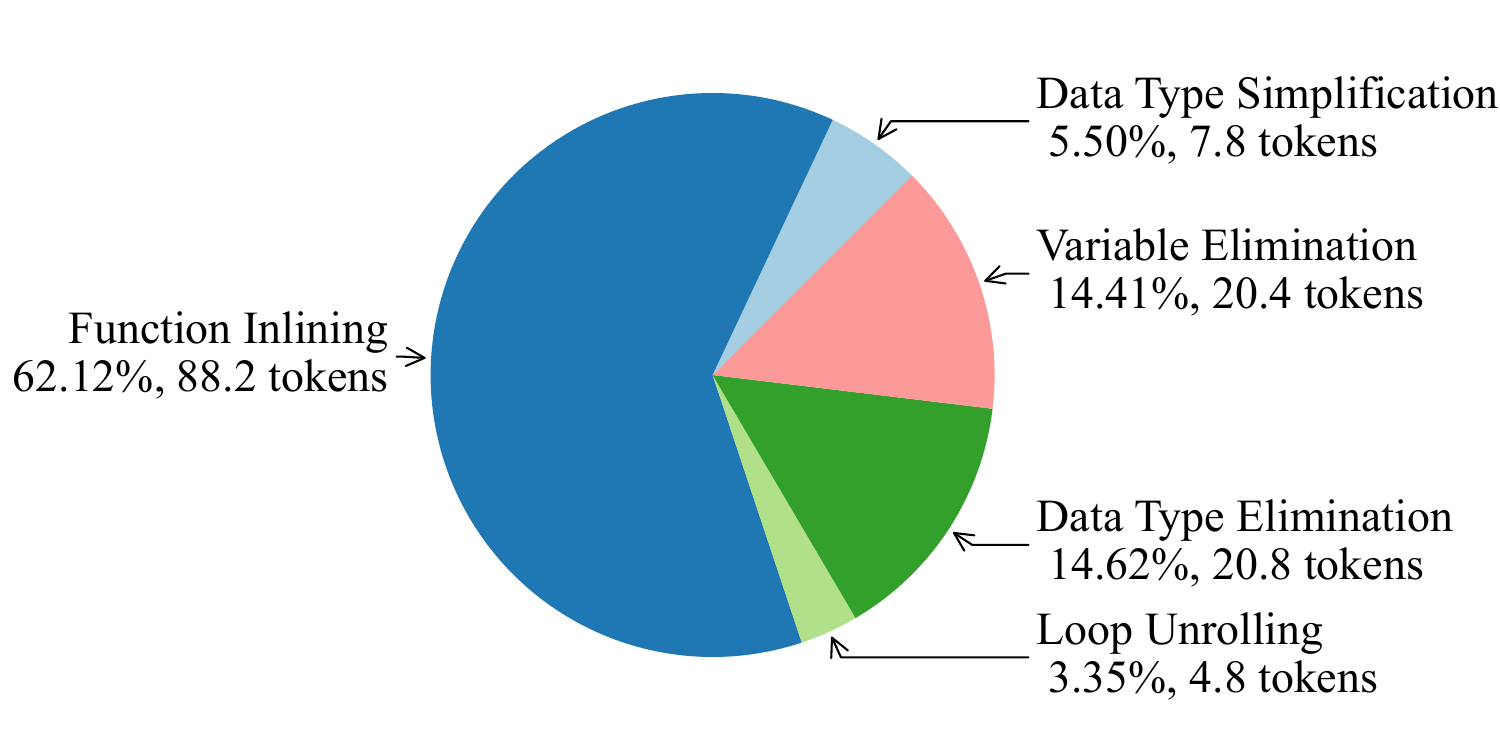}
        \caption{Average size decrease and its percentage induced by each transformation within \benchmarkC.}
        \label{fig:size_changes_piechart}
\end{figure}

\begin{figure}
        \centering
        \includegraphics[width=\linewidth]{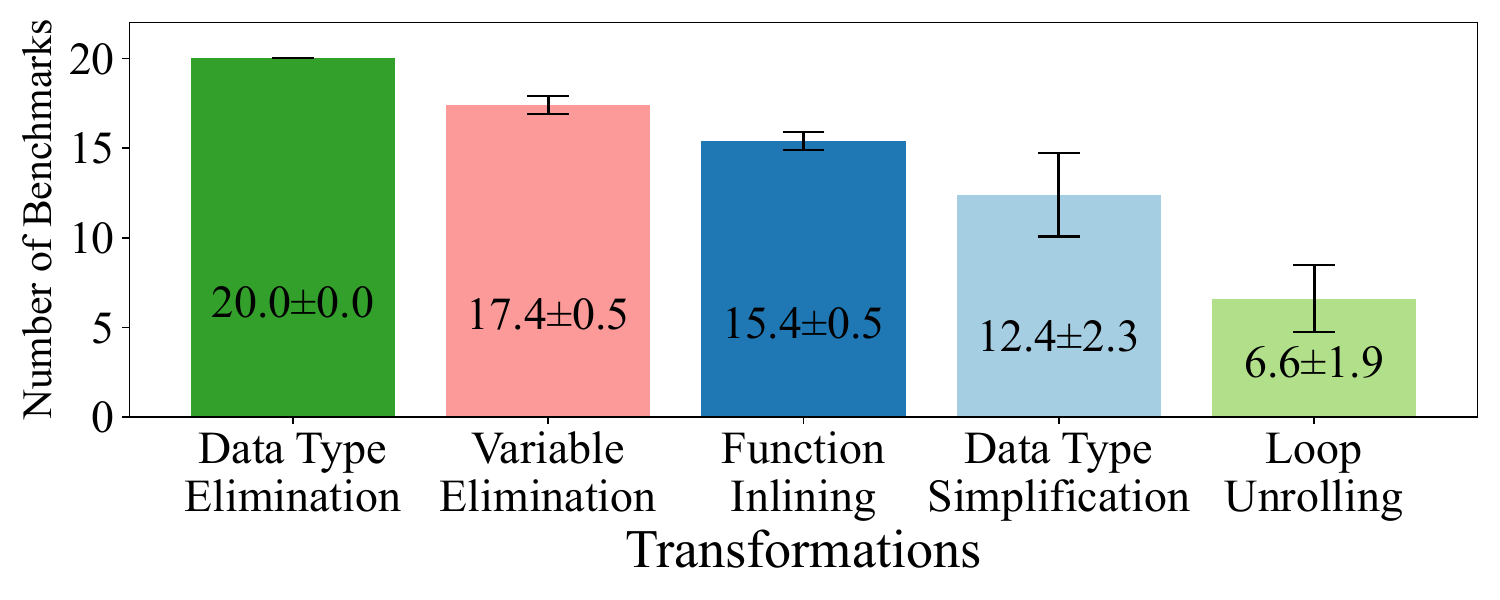}
        \caption{The number of benchmarks impacted by each transformation within
        \benchmarkC.}
        \label{fig:size_changes_barchart}
\end{figure}

%% file: discussion.tex
In this section, we discuss  the effectiveness of multi-level
prompts the
performance
of the \llm under different temperatures, and the failures in \proj.

\subsection{The Effectiveness of Multi-level Prompt}
To validate the effectiveness of our proposed multi-level prompt, we
design the
corresponding single-level prompt and compare its effectiveness against the
multi-level prompt. In detail, different from multi-level prompt, single-level prompt
merges the \primaryQuestionMath and \followupQuestionMath into a single prompt,
\eg, ``Given the following program \{ PROGRAM \}, identify one function
that can be
inlined, and inline it.'' for \opFunctioninlining.

In 5 repeated experiments on  \benchmarkC, the single-level
prompting approach results in $\cSizeAvgSingleLevelProj \pm
\cSizeStddevSingleLevelProj$ tokens, which is far less effective than
results from the multi-level
prompting approach, \ie, $\cSizeAvgProj \pm
\cSizeStddevProj$ tokens.
Our explanation is that, even though such a single-level prompt is
more compact, it makes \llms less concentrated on a specific target, and thus \llms
may omit some targets.

\subsection{The Impact of Temperature}
\label{temperature}
In our experiments, we consistently set the temperature parameter of \llms
to 1.0.
To measure how this parameter affects the performance, we rerun
experiments under multiple temperatures, \ie, 0.75, 0.5,
0.25, 0. Note that higher temperature instructs the \llm to generate more creative
and diverse results. Due to limited resources and time, we evaluate on 10
benchmarks in \benchmarkC with the fastest completion times under
the default configuration.

\input{tables/temperature}
As shown in ~\cref{table:temperature}, the performances under most of the
temperatures are similar, while the exception is \mycode{t=0}, with the average
size worse than others. According to the documentation of
temperature~\cite{openaiapiTemperature},  \mycode{t=0} will actually use a small
threshold above 0. Our speculation is that a low temperature restrains the
diversity of outputs, impeding \proj exploring local minima in different runs,
which is helpful in program reduction tasks.

Given that program reduction is an NP-complete problem, randomness of \llms has
its advantages and drawbacks. Take the results of RQ1 in \cref{table:effectiveness} as
an example.
On the one hand,
randomness allows \proj to explore more distinct local minima, and sometimes
generates smaller programs than \creduce in five repeated
experiments on each benchmark.
On the other hand, randomness of \llms introduces variability. While the standard
deviation remains below 10 in most benchmarks, it may significantly increase in
certain benchmarks, such as $>60$ in \gcc-59903. This variation can be
attributed to the high complexity of the given program. In such scenarios,
\llms
might not
consistently execute accurate transformations, resulting in a range of local minima
and divergent results.

\subsection{Failures in \proj}
In \proj , the generation of variants failing to pass the property test
is common and acceptable. There are two scenarios where \proj fails to
produce variants passing the property test. First, the \llm is
generally
non-deterministic, it may fail to
perform the correct transformation when the program is complex.
Therefore, multiple responses are requested in each \llm query to mitigate
the impact of potential failures.
In another scenario, the transformation may eliminate the bug-triggering
pattern,
e.g., the compiler crashes on a function call, but the transformation inlines
this function. Even if such a transformation is semantically correct,
the property
test will still fail. When the transformed program is bound to fail,
\proj will proceed with the next step using the original program.

Failures indeed lead to more property tests before making any progress,
but they are inevitable, as program reduction is a trial-and-error process.
Even without \llm, property checks issued by \perses and \vulcan usually
have a
failure rate of around 90\%. As shown in ~\cref{table:effectiveness} and
~\cref{table:efficiency}, \proj is
effective and efficient enough, illustrating that failures can be mitigated
via multiple responses.

%% file: tables/temperature.tex
\begin{table}[h!]
	\centering
	\caption{The impact of temperature on reduction sizes.}
	\label{table:temperature}
\resizebox{\linewidth}{!}{%
\setlength{\tabcolsep}{2pt}
\begin{tabular}{@{}l|cc|ccccc@{}}
	\toprule
	\multirow{2}{*}{}           & \multirow{2}{*}{\perses} &
	\multirow{2}{*}{\vulcan} &
	\multicolumn{5}{c}{\proj} \\
	&                          &                          & $t=1$   & $t=0.75$ & $t=0.5$ &
	$t=0.25$ & $t=0$  \\ \midrule
	\multicolumn{1}{c|}{Mean}   & 161.4                    & 102.8                    &
	$73.2\pm3.0$  & $72.2\pm5.0$   &
	$69.5\pm1.0$  & $71.1\pm5.9$   & $90.3\pm3.8$ \\
    \bottomrule
\end{tabular}
}
\end{table}

%% file: threats.tex
In this section, we discuss potential factors that may undermine the validity of our
experimental results.

\subsubsection{Threats to Internal Validity}
The main internal threat comes from the potential data leakage problem.
That is,
do \llms provide reasonable transformation through step-by-step analysis, or just
simply memorize the minimal programs for the benchmark suites, which may be
publicly available on the internet?

We mitigate this threat from several perspectives.
First, program reduction is a task involving programs distinct from those used
in program repair or program synthesis tasks. %
For instance, \llms can learn from large datasets about code generation. %
On the contrary,
programs in program reduction tasks are generally large, complex, and most
importantly, randomly generated to trigger compiler bugs, with
no other specific purpose. Their random and chaotic characteristics make it highly
 unlikely for \llms to memorize such disorganized content, thus reducing the risk of
 data leakage.
Moreover, even
in scenarios where the
\llms might coincidentally memorize certain minimal programs, it is improbable for it
to
link a random-looking, featureless code snippet with a specific memorized program,
especially when no explicit bug ID is provided.
Furthermore, \benchmarkJs, one of the benchmark suites used, was created using
JIT fuzzing tools by the authors and is not publicly accessible. This exclusivity
ensures that the \llms' performance on these benchmarks reflects their ability to
handle unseen and novel programs, thereby showcasing their effectiveness in
managing new challenges without relying on memorized data. This approach
significantly mitigates the risk of data leakage and demonstrates the capacity of \llms
for genuine problem-solving and analysis.

\subsubsection{Threats to External Validity}
One threat to external validity is the generalizability of \proj across languages.
Although the approach of \proj is language-agnostic,
\llms used by \proj may have limited knowledge of
certain languages,
which may affect the performance of \proj.
To mitigate this threat,
we evaluate \proj on three prevalent programming languages, namely, C,
\rust and
\javascript.
The evaluation results demonstrated
the generalizability of \proj on
diverse popular programming languages.
For languages that are not familiar to \llms,
\proj may still produce reasonable results,
since it is shown that \llms like Codex still perform well in
less popular languages like Lua (0.2\% in
Github)~\cite{cassano2022multipl}.
As for a completely new language that \llms cannot recognize and
process,
a possible solution is to incorporate the prompts in \proj with few-shot
prompting~\cite{ahmed2022few, liu2023pre},
so that
\llms learn how to recognize and process a new language from
the given examples.
Further exploration of this approach will be left as future work.

An additional threat is the applicability of our approach across different
\llms. To
mitigate this threat, on the same benchmarks used in ~\cref{temperature},
we repeat the
experiments with \llama, another
widely-used \llm family. As illustrated in ~\cref{table:generality}, the
experimental results on two models of \llama, \ie, \llamaVersiona and
\llamaVersionb,
show no
significant differences from results on \chatgpt.
Furthermore, as \llms continue to evolve, we anticipate improvements in
both quality and time of code processing.

\input{tables/generality}

%% file: tables/generality.tex
\begin{table}[h!]
	\centering
	\caption{The reduction sizes of \proj using various \llms.
	}
	\label{table:generality}
	\resizebox{\linewidth}{!}{%
		\setlength{\tabcolsep}{2pt}
		\begin{tabular}{@{}l|cc|c|cc@{}}
			\toprule
			\multirow{2}{*}{}         & \multirow{2}{*}{\perses} &
			\multirow{2}{*}{\vulcan} & \proj with &
			\multicolumn{2}{c}{\proj with \llama} \\
			&                          &                          &                    \gptCurrentVersion                 &
			13b             &
			34b             \\ \midrule
			\multicolumn{1}{c|}{Mean} & 161.4                    & 102.8                    &
			73.2
			± 3.0                          & 73.4 ± 3.5      & 71.3 ± 4.9      \\ \bottomrule
		\end{tabular}
	}
\end{table}

%% file: related_work.tex
We introduce related work in two topics: program reduction and \llms for
software
engineering.

\subsection{Program Reduction}
\ddmin~\cite{zeller2002simplifying}
initiated the research topic of program reduction.
It treats the input as a list of elements, and consistently splits
the list into halves.
Then it iteratively attempts to reduce the input list by exploring subsets
and their complements at varying levels of granularity, transitioning from coarse to
fine.
\hierarchicaldeltadebugging~\cite{misherghi2006hdd}, short for \hdd, parses
the program input into a
parsing tree, and performs \ddmin on each level of the tree structure.
\perses~\cite{sun2018perses} avoids the generation of syntactically
invalid program variants during reduction by formal syntax transformation.
\vulcan,
further pushes the limit of \perses via identifier/sub-tree replacement and local
exhaustive search~\cite{xu2023pushing}.
RCC is a compact refreshable caching scheme to speed up
program reduction~\cite{rcc}.
 All the above works are not customized for certain
languages, though having high generality, they lack semantic knowledge of a certain
language for further reduction.

Besides, some tools are specifically designed for certain languages.
\creduce~\cite{regehr2012test}, incorporating various transformation passes for
features in \candcpp, is the most effective reducer on this language.
\jreduce~\cite{kalhauge2019binary, kalhauge2021logical} is a tool for \java
bytecode reduction, it reformulates the bytecode reduction into
dependency graph simplification. \ddsmt~\cite{niemetz2013ddsmt} is
designed for reducing programs in \smt format. All these works leverage
language
features to reduce more effectively than language-generic tools.

Distinct from prior work, \proj synergistically combines \llms and
language-generic reduction
tools to harness the advantages of both. Language-generic reducers stand out
for
their remarkable
generality across multiple languages, while \llms excel in
further
refining the programs with the domain knowledge of certain languages learned
from large training sets. Language-specific tools typically demand considerable
human effort to design and implement feature-related transformations for
reduction, while \proj requires only a few lines of natural language prompts,
significantly reducing the effort involved.

\subsection{\llms for Software Engineering}

\llmsfull (\llms) have proved their remarkable capability of undertaking
multiple text-processing tasks, including source
code-related works.
Recent works focus on applying
\llms to facilitate software engineering tasks, or assessing the
effectiveness, potential and limitations of \llms on software development and
maintenance.
Some research~\cite{huang2023empirical, xia2023keep,
xia2023automated, xia2023revisiting} focus on empirically applying \llms
on automatic program repair (\apr).
Huang \etal~\cite{huang2023empirical} performed an empirical
study on improvement brought by model fine-tuning in \apr. Xia
\etal~\cite{xia2023automated} thoroughly evaluated 9 state-of-the-art
\llms across multiple datasets and programming languages, and
demonstrated that directly applying \llms has already significantly
outperformed all existing \apr techniques.
Additionally, some works focus on \llms' performance \wrt code completion,
generation and fuzzing~\cite{liu2023your, zhong2023study,
        tian2023chatgpt, deng2023largezeroshot}, by
leveraging the code analysis and generation ability of \llms.

Similar to these studies, our approach \proj
leverages \llms for
a software engineering task, \ie, program reduction. \proj
harnesses the comprehension and generation capabilities of \llms to
refine the results of program reduction. However, our work
distinguishes itself in the nature of the programs processed by \llms. In
related research, programs are typically logical and goal-oriented, often
designed to fulfill a specific purpose. In contrast, the programs involved in
our program reduction task are random, chaotic, and lack a clear objective.
Consequently, our research sheds light on the performance of \llms when
dealing with programs that do not have an easily discernible purpose.

%% file: conclusion.tex
This paper proposes LLMs-aided program reduction (\proj),
which is the first approach that leverages LLMs for the program reduction task
to the best of our knowledge.
By combining the strength of LLMs and existing language-generic program reduction techniques,
\proj can perform language-specific transformations to effectively reduce the program
while being language-generic (\ie, can be easily applied to a wide range of languages).
The evaluation shows that
in \benchmarkSizeAll benchmarks across three programming languages.
\proj significantly outperforms \vulcan.
Specifically, \proj produces \cSizeDecRateProjVsVulcan, \rustSizeDecRateProjVsVulcan,
and \jsSizeDecRateProjVsVulcan smaller
programs on C, Rust, and
JavaScript, respectively.
Meanwhile, The evaluation also demonstrates that \proj complements \vulcan to some extent.
By reducing the outputs of \proj with \vulcan, we attained results that have similar sizes to those of \creduce
in \benchmarkC.
In terms of efficiency perceived by users, \proj excels in reducing complex programs, and takes
\cTimeDecPercentageOfAvgProjVsVulcan,
\rustTimeDecPercentageOfAvgProjVsVulcan,
\jsTimeDecPercentageOfAvgProjVsVulcan less time than \vulcan
to finish all the benchmarks, respectively.

%% file: data_availability.tex
For replication, we have implemented \proj and released
it publicly~\cite{zhang2024lprartifact}.

%% file: acknowledgments.tex
\begin{acks}
	We thank all the anonymous reviewers in ISSTA'24 for their insightful
	feedback and comments.
	This research is partially supported by
	the Natural Sciences and Engineering Research Council of Canada
	(NSERC) through the
	Discovery Grant, a project under WHJIL,
	and CFI-JELF Project \#40736.
\end{acks}